# Velocity fluctuation and force scaling during driven polymer transport through a nanopore

Martin Charron, Breeana Elliott, Nada Kerrouri, Liqun He, Vincent Tabard-Cossa*

150 Louis-Pasteur Private, Department of Physics, University of Ottawa, Ottawa K1N 6N5, Canada

*Corresponding Author: tcossa@uottawa.ca

**Abstract**

Inspired by its central role in many biological processes, the transport of biopolymers across nanoscale pores is at the heart of a single-molecule sensing technology aimed at nucleic acid and protein sequencing, as well as biomarker detection. When electrophoretically driven through a pore by an electric potential gradient, a translocating polymer hinders the flow of ions, producing a transient current blockage signature that can be mapped to physicochemical properties of the polymer. Although investigated theoretically and by simulations, few experimental studies have attempted to validate the predicted transport properties, mainly due to the complex nature of the non-equilibrium translocation process. Here, we elucidate these fundamental concepts by constructing a patterned DNA nanostructure whose current signatures allow measurement of the instantaneous velocity throughout the translocation process. With simple physical insights from polymer and fluid dynamics, we show how the resulting molecular velocity profiles can be used to investigate the nanoscale forces at play and their dependence on experimental parameters such as polymer length, pore size and voltage. These results allow testing of theoretical models and outline their limitations. In addition to bridging experiment and theory, knowledge of the velocity fluctuation and force scaling during passage can assist researchers in designing nanopore experiments with optimized sensing performance.

The passage of polymers through pores is a ubiquitous process observed in cellular systems, often driven by pH or chemical potential gradients. When driven by an electric potential gradient, the electrophoretic passage of polymers can be identified and characterized by monitoring transient ionic current blockages due to the polymers impeding the flow of ions through the channel.[1] This principle is at the core of many highly successful applications and exciting research endeavors including nucleic acid sequencing, biomarker detection, and more recently protein fingerprinting and synthetic polymer decoding.[2–5] Although studied theoretically and through simulations,[6–14] the physics of driven polymer translocations has yet to be extensively characterized experimentally, mostly due to its highly non-equilibrium nature: Under voltages commonly used for sensing (50 – 1000 mV), translocation times of DNA have been consistently reported to be significantly smaller than their corresponding relaxation times.[15] At any instant during the process, the polymer is thus in a conformation that differs significantly from ones adopted when relaxed in bulk solution.

Thus far, the kinetics of electrophoretically driven polymer passages through nanopores have been studied experimentally by reporting the dependence of polymer translocation durations on the applied voltage, polymer length, pore dimension, and other experimental parameters such as bulk salt concentration and viscosity.[16–22] Additionally, electrohydrodynamic forces in nanopore systems have been measured through the use of optical tweezers inserting and stalling DNA inside nanopores under an applied voltage[23–27], or more recently through the use of DNA-origami sphere docked atop nanopores trapping polymers inside pores.[28–30] Although rich in information, such measurements of the total translocation durations and forces under fixed polymer conformations fail to elucidate the time-dependent forces and velocities expected from the non-equilibrium process described by theory and simulations,[6–14] vital information for applications like scaffold-assisted sensing that attempt to locate features along the backbone of a DNA carrier based on temporal signals. To this end, recent experimental works have made use of nanostructured DNA molecules to estimate the instantaneous velocity throughout the translocation process.[15,31,32] In particular, Chen *et al*. showed that driven polymer translocations are a two-step process, wherein polymers initially slow down before accelerating towards the end of their passage,[15] in qualitative agreement with the principles of Tension Propagation introduced by Saito and Sakaue[6].

In this work, we go beyond the qualitative observation of the translocation velocity profiles of previous experimental studies,[15,31,32] and provide a detailed quantitative report on the dependence on experimental parameters, including polymer length $L$, pore diameter $d_{pore}$ and applied voltage $\Delta V$, of high practical value for the operation of many sensing schemes. These measurements employ DNA nanostructures with interspaced domains of different cross-sectional areas, allowing precise experimental estimates of the instantaneous translocation velocity. In addition to demonstrating the origin of scalings between translocation time and $L$, $d_{pore}$ and $\Delta V$, our results characterize the underlying time-dependent forces imparted on translocating polymers. Comparison of our experimentally observed trends with predictions from Tension Propagation and fluid dynamics show good agreement, while also outlining shortcomings and limitations of these theoretical concepts. In parallel to bridging theory and experiment, by characterizing different metrics of velocity profiles we are able to provide answers to simple practical questions such as: How much does velocity fluctuate during translocation? Which polymer segment is inside the pore when the velocity is minimal? This knowledge provides great insights for anyone designing nanopore experiments with optimized sensing performance, whether to select the optimal pore dimensions, or for the architecture of a molecular probe or carrier. Finally, we also report a non-intuitive finding that translocation times increase in larger pores.

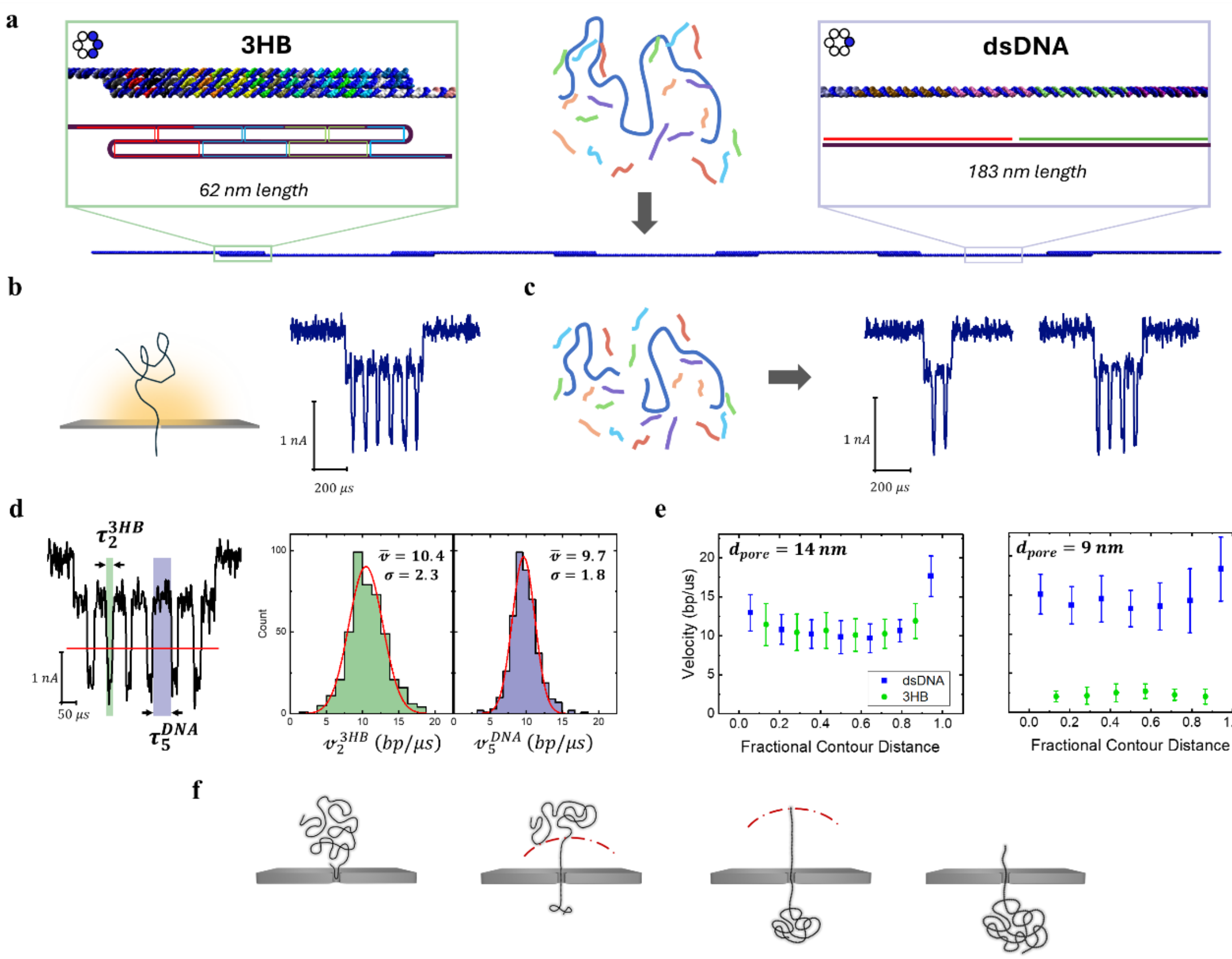

**Figure 1. a)** Design for the Velocity Profiling Nanostructure (VPN), consisting of six 3-Helix Bundle (3HB) segments interspaced by seven dsDNA segments **b)** Signals of single-file passage of fully assembled VPN **c)** Signals of single-file translocations of partially assembled VPN. **d)** Demonstration of threshold-crossing algorithm used to obtain segment duration and velocity statistics. Velocity distributions of a 3HB and DNA segment are shown with corresponding gaussian fits, obtained from passages through a 14 nm pore. **e)** Translocation velocity profiles obtained from a 14 nm pore, and a 9 nm pore. Error bars correspond to the standard deviation of the fitted velocity distributions. **e)** Sketch of Tension Propagation (TP) occurring during translocation.

## Nanostructure Design and Velocity Measurements

A patterned DNA nanostructure was designed with two differently sized repeating segments such that the current trace resulting from its passage through a nanopore would allow the estimation of the polymer's instantaneous translocation velocity. The Velocity Profiling Molecule (VPM) was assembled by mixing a 7292 nucleotide (nt) long single-stranded DNA (ssDNA) scaffold (m13 phage) with 171 short ssDNA staples, resulting in a linear structure with thirteen interspaced domains: seven double-stranded (dsDNA) segments and six three-Helix Bundles (3HB) segments, the latter corresponding to the scaffold folded twice on itself, and thus to three dsDNA segments in parallel.[33,34] See Methods and section S1 of the SI for more details on the assembly protocol, sequences, and exact segment dimensions. Due to their different cross-sectional areas and thus different induced current blockages, the passage of individual segments can be identified in the current trace produced by the single-file passage of a VPM through a nanopore (Figure 1b). As expected, the blockage amplitudes from 3HB segments $\Delta I_{3HB}$ are consistently observed to be three times deeper than those from the dsDNA spacers $\Delta I_{DNA}$. The end-to-end length of VPMs was designed to be nearly exactly that of 5 kbp dsDNA. As such, 5 kbp dsDNA was passed through pores before VPMs as a control to ensure similar total translocation durations, thus confirming that the presence of 3HB segments did not appreciably affect the translocation dynamics. Note that we took advantage of the presence of fragmented m13 scaffolds in the assembly mixture resulting in VPMs of different lengths (Figure 1c) to study the effect of polymer length on velocity profiles on the same pore. We also note that only single-file VPM translocations were analyzed for this work, although events where the polymer entered the pore not by an extremity but instead by bending somewhere along its contour length and folding inside, were commonly observed.[35]

To infer translocation velocities from individual event traces, a simple threshold-crossing algorithm was used, wherein a blockage threshold was set to a value between $2\Delta I_{DNA}$ and $3\Delta I_{DNA}$ away from the baseline, and times at which the current trace crosses the threshold either upwards or downwards were noted (Figure 1d). The passage duration $\tau_i$ of the $i^{th}$ segment of a VPM could then be determined as the time interval between corresponding threshold crossing times of the $i^{th}$ sub-level in the current trace, with the event start and end times of the entire event used to delimit the first and last segments, respectively. Given the known length $\ell_i$ of the $i^{th}$ segment, its translocation velocity was calculated as $\mathcal{v}_i = \ell_i/\tau_i$, with units of $bp/\mu s$. Figure 1d shows the histogram of the velocities measured from multiple individual VPM translocations for the second 3HB and fifth dsDNA segments, $\mathcal{v}_2^{3HB}$ and $\mathcal{v}_5^{DNA}$, measured in a 14 nm nanopore in 3.6 M LiCl under an applied voltage of 200 mV. By fitting each segment's velocity distribution to a Gaussian function, the mean velocity $\overline{\mathcal{v}_i}$ and the standard deviation $\sigma_i$ of the $i^{\text{th}}$ segment velocity can be extracted. Figure 1e plots the mean translocation velocities calculated for each segment versus the segment center's location relative to the VPM's contour for a 14 nm and a 9 nm diameter nanopore (± 1 nm). The magnitude of the velocity error bars corresponds to the segments' extracted $\sigma$ values. Section S2 of the SI discusses the robustness of the analysis approach through its sensitivity to fitting parameters and methods.

Qualitatively, Figure 1e shows that the calculated velocity profiles are non-uniform, non-monotonic, and consistent with prior experimental work:[15] After translocation begins from the *cis* side, the polymer decelerates until roughly its contour midpoint, after which it speeds up until fully exiting the pore on the *trans* side. Such velocity profiles are expected from tension propagation principles[6–11] stating that a polymer is not in equilibrium throughout its translocation process, and as such only monomers under tension are in motion and impart a hydrodynamic drag opposing the electrophoretic pulling force. Initially, as translocation progresses, the number of monomers under motion increases as the tension front progresses along the polymer contour. Once the tension front has reached the back end of the polymer however, its tail-end starts retracting towards the pore, and as such the number of monomers in motion starts to reduce over time.

The velocity profile for the 9 nm pore (Figure 1e) reveals the importance of pore-polymer interactions and the quasi-static nature of the translocation process by showing 3HB segments moving significantly slower than dsDNA segments. This is not surprising, since their different cross-sectional areas results in 3HB segments interacting more strongly with the pore walls than the smaller DNA segments, and pores with similar diameters to translocating polymers are known to significantly slow down translocations times due to steric interactions.[18] Most pores used in this work are larger than 9 nm, and show similar velocity profiles for 3HB and dsDNA segments, as expected since both polymers were previously measured to have similar electrophoretic mobilities.[33] For readability, most velocity profiles plotted throughout this text do not include error bars (see section S3 of the SI for profiles with error bars).

## Polymer Length Dependence

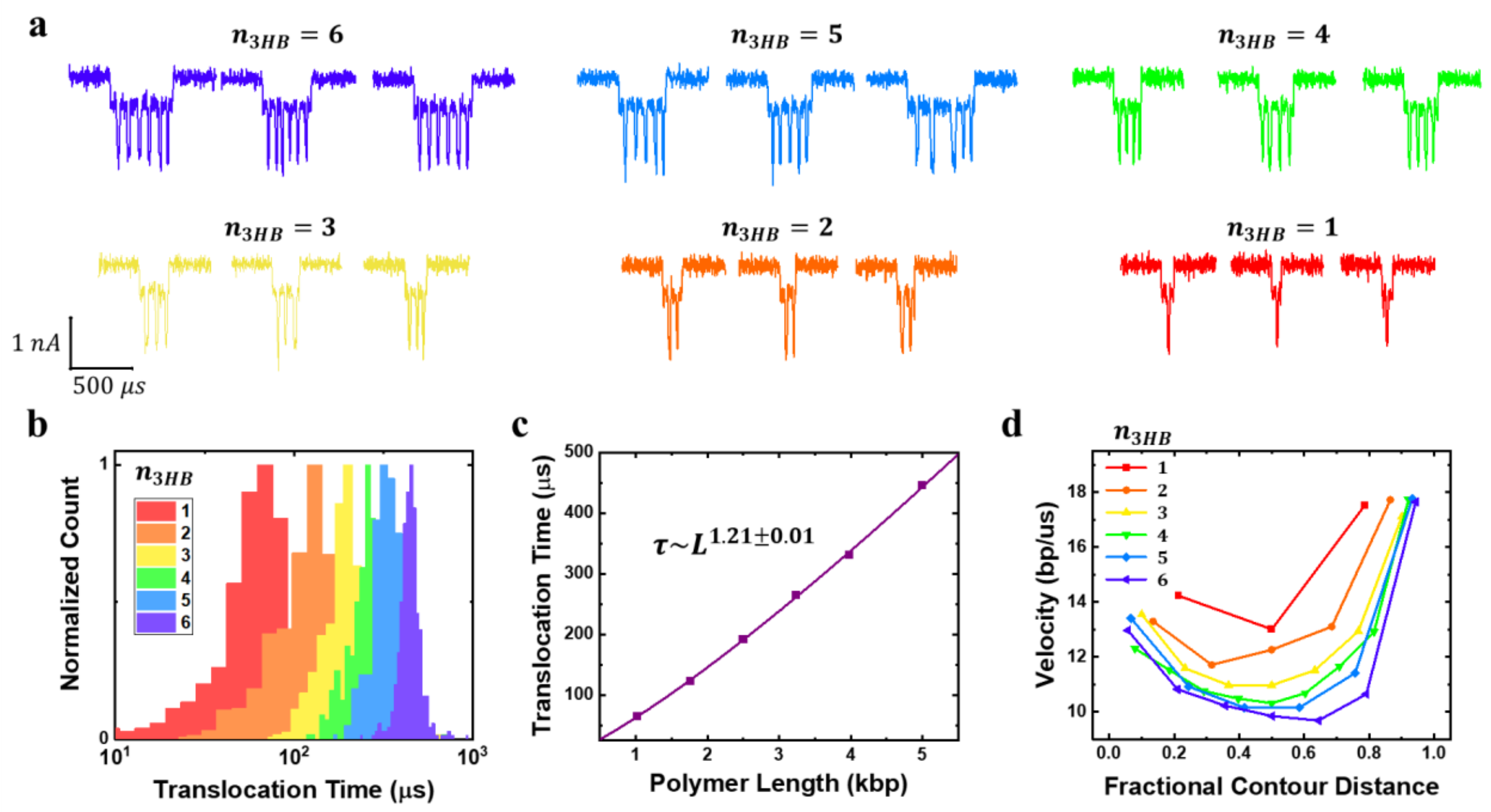

**Figure 2.** Effect of polymer length on translocation through a 14 nm SiN pore under a 200 mV bias in 3.6 M LiCl, 10 mM HEPES pH 8. **a)** Current traces of six different VPM lengths. Traces are classified by $n_{3HB}$, i.e. the number of detected 3HB segments. Data recorded at 4.16 MHz sampling rate and low-pass filtered at 300 kHz for analysis. **b)** Translocation time distributions of VPMs of different lengths. **c)** Dependence of mean translocation time on VPM length $L_n$ (Eq. 1). **d)** Extracted mean velocity profiles. Bump velocities are included for $n_{3HB} \leq 4$ for better spatial resolution.

We now describe the effects of polymer length on translocation velocity profiles measured using single-file translocations of VPMs of different lengths. Figure 2a shows current traces of VPM translocations through a 14 nm pore under a 200 mV bias in a 3.6M LiCl solution, wherein individual translocation events were classified according to the number of 3HB segments $n_{3HB}$ detected per current trace. Figure 2b plots the normalized distributions of translocation times for each VPM population. Distributions were fit to log-normal functions from which the mean translocation time $\bar{\tau}_n$ of each population was obtained. Figure 2c shows the dependence of mean translocation time $\bar{\tau}_n$ on the effective polymer length $L_n$ of each length of VPMs, which were calculated by assuming a single scaffold fragmentation occurred with equal probability along the scaffold contour, and thus along the dsDNA segments. For example, the average VPM length for $n_{3HB}$ 3HB segments is thus:

$$L_n = (n_{3HB} + 0.5)L_{dsDNA} + n_{3HB}L_{3HB} \tag{1}$$

where lengths $L_{DNA}$ and $L_{3HB}$ are the lengths of the dsDNA and 3HB segments, respectively. The data of Figure 2c is well fitted by a power scaling function: $\bar{\tau}_n \propto L_n^{\alpha}$ with a power scaling coefficient of $\alpha = 1.21 \pm 0.01$. Such power scalings are commonly observed with dsDNA translocations through nanopores, with experimental coefficients ranging in value between 1.19 to 1.4.[17,20,22,36,37] We note that the 1.21 coefficient from VPMs is identical to the coefficients reported in prior publications for dsDNA measured on SiN nanopores fabricated by the control breakdown method.[20,36,38] This supports our hypothesis that these structured VPMs are adequate proxy molecules to study the effect of polymer lengths on translocation dynamics of dsDNA polymers.

Figure 2d plots the velocity profiles for each VPM length. All profiles share a similar two-step non-monotonic shape, as in Figure 1e. Interestingly, the initial and final velocities, $v_{ini}$ and $v_{end}$, appear to be independent of polymer length as they are consistently measured to be $v_{ini} = 13.3\ bp/\mu s \pm 0.6\ bp/\mu s$ and $v_{end} = 17.6\ bp/\mu s \pm 0.2\ bp/\mu s$. The velocity profiles of the different

VPM lengths however differ in how much they fluctuate throughout the passage: The minimal velocity achieved $v_{min}$ is consistently measured to be lower for longer polymers. Moreover, the fractional location where the minimal velocity occurs, $x_{min}$, is consistently lower for shorter polymers. Without employing any physical insights yet, Figure 2d empirically demonstrates why the dependence of translocation time on polymer length shows a super-linear power scaling ($\tau \sim L^{1.21}$ in this work). Although translocations of different polymer lengths begin and end at the same velocity, longer polymers simply slow down more importantly throughout the process than shorter ones. This is in contrast with a linear dependence, $\tau \sim L$, which would result in length-independent velocity profiles with identical $v_{in}$, $v_{min}$ and $v_{end}$.

**Pore Size Dependence**

We now describe the effects of pore size on translocation velocity profiles. Figure 3a shows current traces obtained from the passage of VPMs in nanopores of diameters ranging between 9 nm and 36 nm, measured under a 200 mV electrical bias and in a 3.6 M LiCl solution. Figure 3b shows the extracted mean translocation velocity profiles. Velocity profiles in Figure 3b show a clear dependence on pore size: the velocities across the profile consistently are reduced with increasing pore size. This is further illustrated in Figure 3c which plots the dependence of the final velocity $v_{end}$ on pore diameter. Moreover, with the spatial resolution provided by the design of the VPM, the contour location of the minimal velocity $x_{min}$ appears to be independent of pore size.

Interestingly, across all the velocity profiles shown in Figure 3b, the differences between the final segment velocity $v_{end}$ and the minimal velocity $v_{min}$ appear to be similar across the different pore diameters with $v_{end} - v_{min}$ values ranging between $6.3\ bp/\mu s$ and $8.5\ bp/\mu s$. Given the consistently lower velocities measured for larger pores, velocity fluctuations are thus more pronounced for larger pores than for smaller ones. For example, the velocity difference $v_{end} - v_{min} = 7.0\ bp/\mu s$ measured in a 36 nm pore corresponds to 59% of $v_{end}$, whereas the difference of 6.3 $bp/\mu s$ in a 9 nm pore represents 30% of $v_{end}$. Figure 3d further exemplifies this by plotting the velocity profiles normalized by the end velocity $\overline{v_i}/v_{end}$, where the normalized value $v_{min}/v_{end}$ is shown to reduce monotonically with increasing pore size. We note the qualitative similarity between Figures 2d and 3d: Increasing pore size or polymer length monotonically reduces the value of $v_{min}/v_{end}$, or equivalently increases the relative velocity fluctuations observed during translocation events.

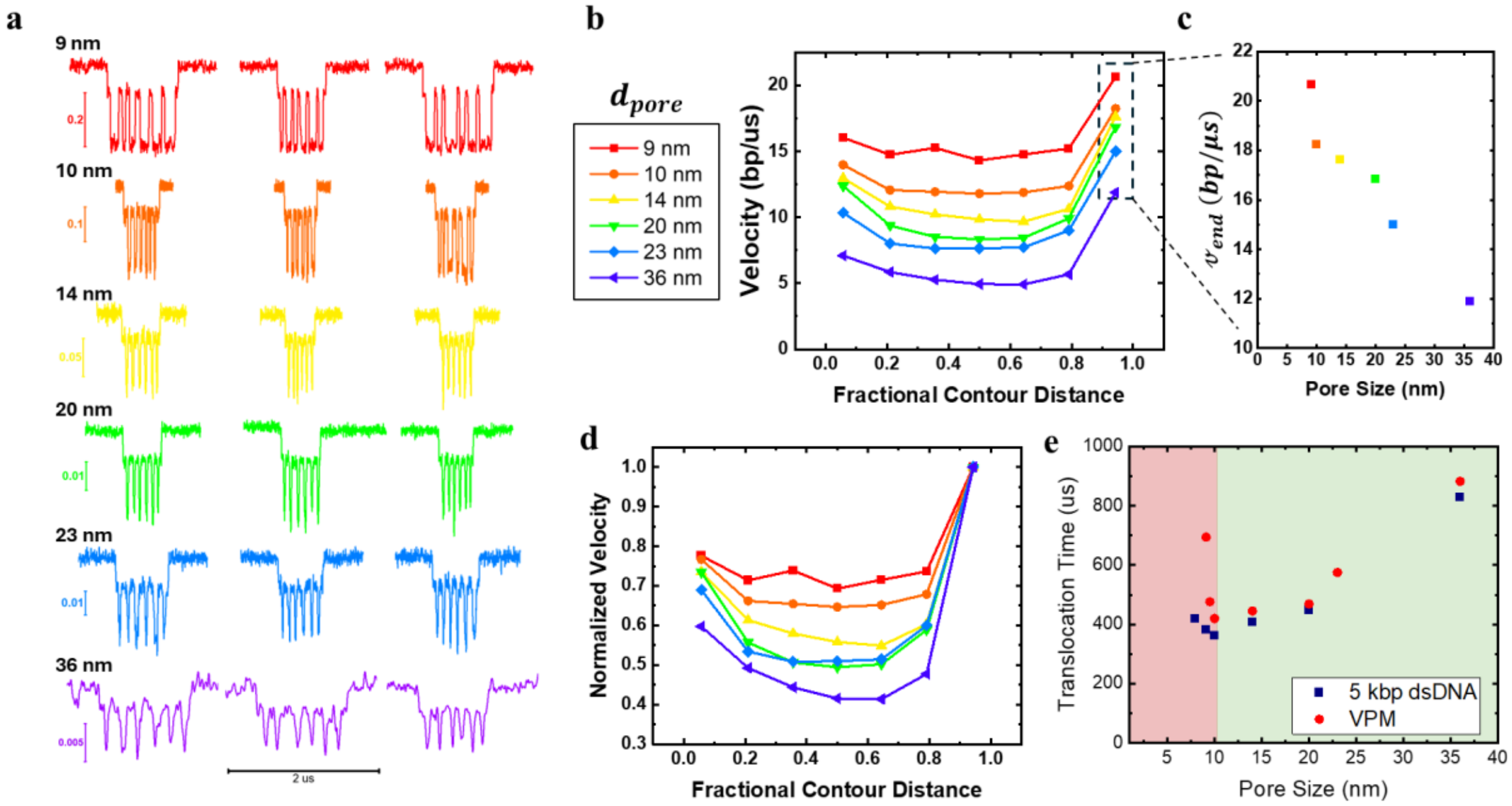

**Figure 3.** Effect of pore size on translocation dynamics under a 200 mV bias in 3.6 M LiCl, 10 mM HEPES pH 8. **a)** Current traces of VPM translocations in six different nanopore diameters. Scale bar values are normalized by open pore current. **b)** VPM velocity profiles for different pore sizes. **c)** Dependence of $v_{end}$ on pore size. **d)** VPM velocity profiles normalized by $v_{end}$. **e)** Translocation times of VPM structures and 5 kbp dsDNA measured in different pore sizes. Red and green colors depict the two size-dependence regimes, as detailed in the text.

Before quantifying and expanding on the dependence of $v_{min}/v_{end}$ on pore size and polymer length, we comment on the unexpected result from Figures 3a-c showing slower translocation velocities for large pores. To our knowledge, except for Figure S2 in the SI of Garaj *et al.*[39] , the translocation times have always been reported as getting faster with increasing pore size until reaching a plateau due to reduced pore-polymer steric interactions.[18,22,40] To further depict this, Figure 3e plots the mean translocation times $\bar{\tau}$ of VPM structures and 5 kbp dsDNA measured in different pore diameters $d_{pore}$, which shows that $\bar{\tau}$ decreases with $d_{pore}$ for small pores, as per previous reports,[18,22,40] whereas $\bar{\tau}$ increases with $d_{pore}$ for larger pore sizes. More examples of similar behavior are shown in section S4 of the SI. Although not studied more in depth, we suggest that the size of the sensing region increases with $d_{pore}$ due to the more important access resistance contributions in larger pores, resulting in longer length travelled between the beginning and end of the current blockade.

**Velocity Fluctuation Toy Model**

With insights from Figures 2 and 3, we now quantitatively study the velocity fluctuations observed during the translocation process. Namely, through the help of a toy model, we use experimental velocity fluctuations to study the origin and scalings of the time-dependent forces at play during the electrophoretically driven pore transport process. To model the forces imparted on translocating polymers, we first separate them into two broadly defined classes: forces arising inside and outside the pore, as sketched in Figure 4a. Importantly, given that the electric field in finite-length pores extends outside the channel, we loosely define the interior (exterior) of the pore as the space where forces arising from the electric field and electrokinetic effects are significant (insignificant).

Inside the pore, as per the stalled DNA measurements,[24] we consider the electric pulling force resulting from the electric field pulling on the charged DNA backbone $-F_e$, and the hydrodynamic drag imparted by the electroosmotic flow resulting from the motion of counterions shielding both the charged pore surface and polymer $F_{eo}$. Additionally, we consider the hydrodynamic drag resulting from the polymer moving at a velocity $\mathscr{v}$ through the pore interior as $\gamma_{in}\mathscr{v}$, where $\gamma_{in}$ is the corresponding internal drag coefficient, which is time-independent throughout translocation. Outside the pore, following tension propagation insights,[6–11] we consider the hydrodynamic drag force resulting from the motion of a segment of length $\ell$ under tension and moving at the same velocity $\mathscr{v}$ as segments inside the pore $\gamma_{out}(\ell)\mathscr{v}$. Here $\gamma_{out}(\ell)$ denotes the corresponding external drag coefficient and is expected to scale linearly with the length of the polymer in motion, i.e. $\gamma_{out}(\ell)\sim\ell$, which according to tension propagation is dependent on time. Note the directions of the forces arise from the assumption of negatively charged polymer and pore walls, and a *trans* to *cis* pointing electric field, as shown in Figure 4a. For a polymer segment of length $\ell$ under tension, the translocation velocity can be simply written:

$$\mathscr{v}(\ell) = \frac{F_e - F_{eo}}{\gamma_{in} + \gamma_{out}(\ell)} = \mathscr{v}_{end}\left(1 + \frac{\gamma_{out}(\ell)}{\gamma_{in}}\right)^{-1} \tag{2}$$

The final expression of Equation 2 is obtained by recognizing that the end velocity simply corresponds to the case where $\ell = 0$, i.e. $\mathscr{v}_{end} = (F_e - F_{eo})/\gamma_{in}$.

According to Equation 2, the toy model suggests that at any instant during the translocation process, the normalized velocity $\mathscr{v}/\mathscr{v}_{end}$ depends solely on the ratio of internal and external drag coefficients $\gamma_{out}(\ell)/\gamma_{in}$. We suggest this explains why Figure 2d (velocity profile versus polymer length) and Figure 3d (velocity profile versus pore size) look alike: Longer polymers result in larger segments under tension outside the pore on average, thereby increasing $\ell$ and $\gamma_{out}(\ell)$, and larger pores result in weaker confinement of the translocating polymer and thus reduce the internal drag, $\gamma_{in}$. Both longer polymers and larger pores therefore result in increased values of $\gamma_{out}(\ell)/\gamma_{in}$ throughout the translocation process, and thus in overall slower translocations. Moreover, experimental conditions wherein the time-independent internal drag $\gamma_{in}$ dominates over the external time-varying drag $\gamma_{out}$, i.e. in the case of small pores and short DNA, should result in flatter profiles with less fluctuations, whereas the opposite is true for conditions where the time-dependent $\gamma_{out}$ dominates over the time-independent $\gamma_{in}$, i.e. the case of long polymers and large pores.

To quantitatively analyze the velocity profiles, we further note that the minimal velocity $v_{min}$ corresponds to the point during translocation where the segment under tension is maximal, i.e. $v_{min} = v(\ell_{max})$. Defining the maximal external drag coefficient as $\gamma_{out}^{max} \equiv \gamma_{out}(\ell_{max})$, it can be shown that:

$$\frac{\gamma_{out}^{max}}{\gamma_{in}} = \frac{v_{end}}{v_{min}} - 1 \tag{3}$$

Equation 3 shows that the experimentally obtained values of $v_{end}/v_{min}$ from Figures 2 and 3 can be used to quantify the maximal external and internal drag coefficient ratio $\gamma_{out}^{max}/\gamma_{in}$. Figure 4d plots the values of $v_{end}/v_{min} - 1$ calculated from the different polymer lengths $L_n$ shown in Figure 2d and shows that $\gamma_{out}^{max}/\gamma_{in}$ increases with polymer length. Namely, the data is well-fitted by a power scaling function of the form $L_n^{0.55\pm0.04}$. Since the internal drag $\gamma_{in}$ is independent of polymer length, and the maximal external drag $\gamma_{out}^{max}$ is proportional to $\ell_{max}$, we can further infer from our experiments that $\ell_{max}$ scales with polymer length as $\ell_{max} \sim L^{0.55\pm0.04}$. From tension propagation principles, under strong pulling forces, $\ell_{max}$ should be closely related to the polymer's radius (Figure 1), which in bulk solution should scale as $R_{ee} \sim L^{\nu}$ with $\nu = 0.588$ being Flory's excluded volume coefficient. Great agreement is thus found between our experimental results and predictions from Tension Propagation. Similar agreement was found using VPM events passing through a 20 nm pore.

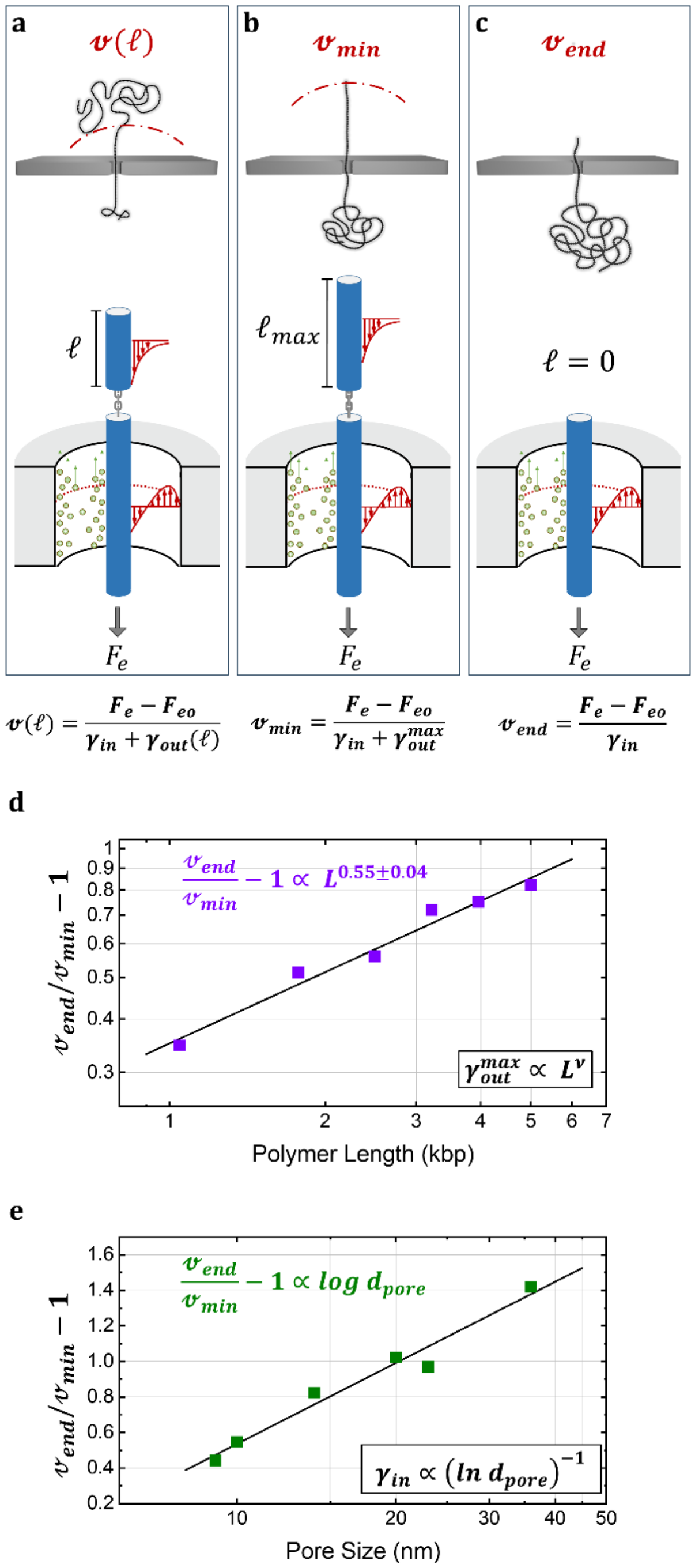

**Figure 4.** Toy model and scaling of velocity fluctuations (Eq. 2). **a-c)** Depiction of forces considered for different translocation steps. **d)** Dependance of $v_{end}/v_{min} - 1$ on polymer length $L_n$ calculated from Eq. 1. The solid line is a power scaling fit resulting in a scaling coefficient of $\gamma_{out}^{max}/\gamma_{in} \sim L^{0.55\pm0.04}$. **e)** Dependance of $v_{end}/v_{min} - 1$ on pore diameter. The solid line is a logarithmic fit of the form $\gamma_{out}^{max}/\gamma_{in} = 0.66\ln(d_{pore}) - 0.98$. Data taken from Figures 2d and 3d.

Figure 4e plots the calculated values of $v_{end}/v_{min} - 1$ obtained from Figure 3d against corresponding pore diameters and shows once again that $v_{end}/v_{min} - 1$ and thus $\gamma_{out}^{max}/\gamma_{in}$ increases with pore sizes. The dependence of $\gamma_{out}^{max}/\gamma_{in}$ on pore size appears to be logarithmic, as supported by the data being well fitted by a function of the form $\gamma_{out}^{max}/\gamma_{in} = A\ln(d_{pore}) - B$, where A and B are constant. By assuming the external coefficient $\gamma_{out}^{max}$ is independent of pore size, we can further infer from our experiments that the internal drag coefficient scales as $\gamma_{in}^{-1} \sim \ln(d_{pore})$. Interestingly, this inverse logarithmic pore size dependence arises when calculating the drag coefficient of an infinitely long cylindrical object moving through an infinitely long cylindrical channel (Section S5 of the SI). Although no analytical solution exists for the internal drag coefficient $\gamma_{in}$ of a finite-length nanopore, our translocation velocity experiments suggest that the pore size scaling appears to be maintained, at least to first order.

**Polymer Conformations**

We now discuss the effects of polymer conformation on velocity profiles. We begin by demonstrating how the fractional location of the segment with minimal velocity $x_{min}$ along the VPM contour can be predicted reasonably well using insights from Tension Propagation. As discussed above, minimal velocity occurs when the segment under tension has reached its maximal length $\ell_{max}$. In the extreme limiting case of ultra-fast translocations, the maximal length $\ell_{max}$ corresponds to the distance between the pore entrance and the polymer segment farthest away from the pore at the onset of translocation, $R_{max}$, as shown in Figure 5a. A reasonable expression for $x_{min}$ is therefore:

$$x_{min} = \frac{L_n - \ell_{max} - \ell_{end}}{L_n} = 1 - \frac{R_{max} + \ell_{end}}{L_n} \tag{4}$$

Here, $\ell_{end}$ simply corresponds to the length of segments not yet under tension when the farthest segment becomes under tension, i.e. once the translocation velocity is minimal. Note that Figure 4b depicts a scenario wherein $\ell_{end} = 0$, as opposed to Figure 5a.

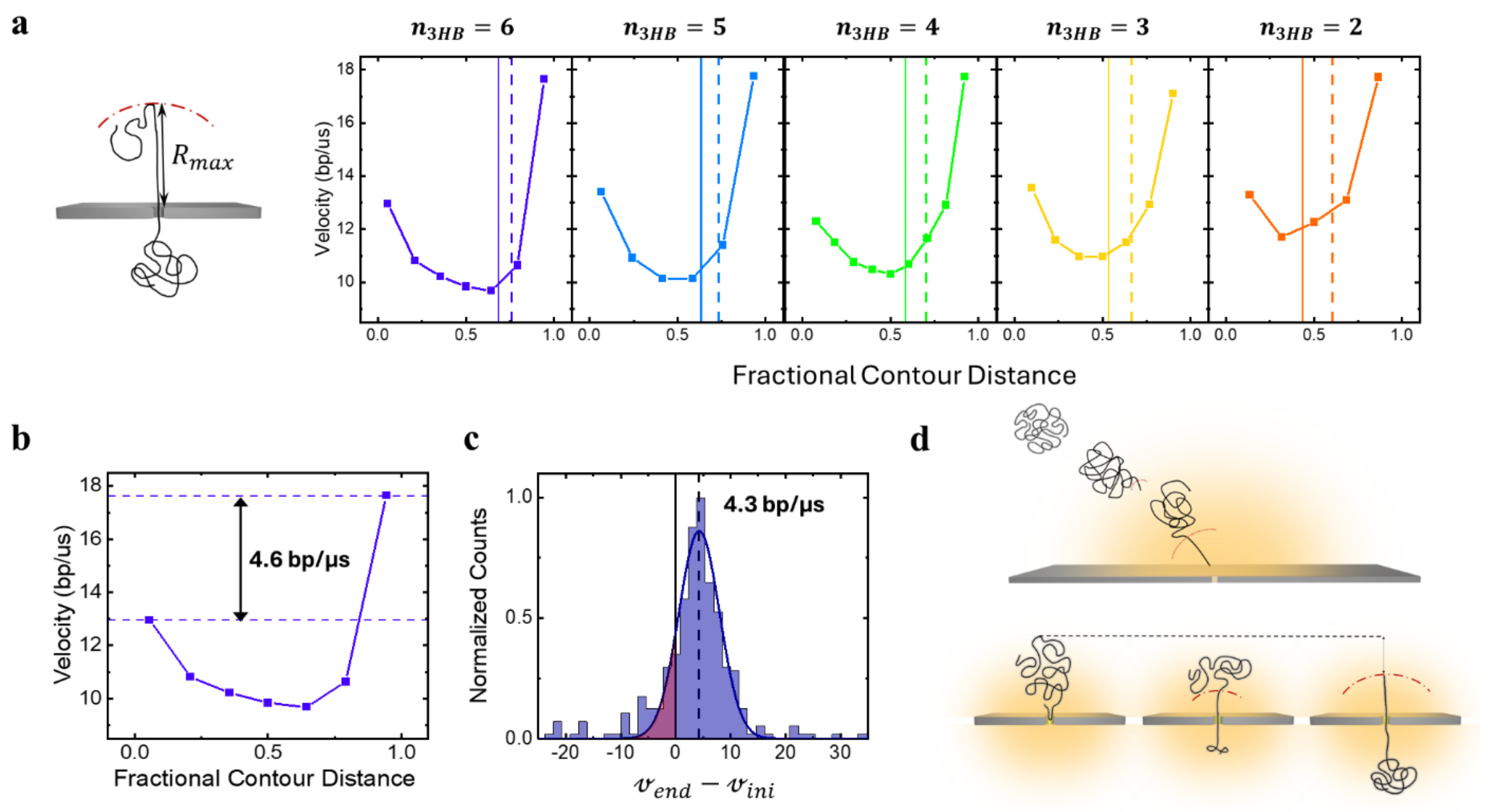

**Figure 5.** Effects of polymer conformations on velocity profiles. **a)** Translocation Velocity Profiles of different VPM lengths. Vertical lines represent $x_{min}$ predictions calculated using $x_{min} = 1 - R_{ee}/L_n$ (dashed) and using most probable $x_{min}$ values extracted from 5000 simulated discrete worm-like chains (solid). **b)** Difference of average start and end velocities. **c)** Distribution of $v_{end} - v_{in}$ values measured for 439 individual events. **d)** Interpretation of the effect of electric field on polymer conformations, as described in the text.

The simplest attempt to solve Equation 4 is to assume that the length of polymer on the *cis*-side when $v = v_{min}$ is equivalent to the polymer's relaxed end-to-end distance: $R_{ee} = R_{max} + \ell_{end}$. Using wormlike chains statistics, $R_{ee}$ values were calculated for different VPM lengths $L_n$ using a persistence lengths of 150 bp. Figure 5a shows the individual translocation velocity profiles for the different VPM lengths (data from Figure 2) and includes the $x_{min} = 1 - R_{ee}/L_n$ predictions plotted as dashed vertical lines, which consistently overestimate the location of $v_{min}$. To better quantify $R_{max} + \ell_{end}$ for the calculation of $x_{min}$, we generated 5000 discrete wormlike chains (DWLC) with one end tethered to a non-penetrable surface, as detailed in section S6 of the SI.[41] By identifying the furthest monomer from the pore (tethering point), statistics on $R_{max} + \ell_{end}$ and $x_{min}$ (Eq. 4) were obtained. Figure 5a plots the most probable simulated $x_{min}$ values as solid vertical lines, which line up well with local velocity minima, thus establishing that the location of minimum velocity is closely correlated to the relaxed conformation of a polymer with one end inside the pore.

For all velocity profiles obtained and shown in this work, the average initial translocation velocities $\bar{v}_{ini}$ were consistently smaller than the average end velocities, $\bar{v}_{end}$. This is revealed in Figures 5b-c which show the difference $\bar{v}_{end} - \bar{v}_{ini} = 4.6\ bp/\mu s$ and the distribution of $v_{end} - v_{ini}$ with 88% of events showing $v_{end} > v_{ini}$. According to the toy model introduced above, $v_{end} > v_{ini}$ suggests the polymer segments closest to the pore are already under tension when translocation begins. This can either arise from conformational bias of analyzing only single-file translocations or from tension propagation beginning prior to entering the sensing region due to the electric field extending outside the pore. The latter is consistent with simulations that show polymers being stretched and elongated by the external non-uniform electric field upon their approach.[12–14,42] During the translocation process, the non-uniform external field was further shown to compress the polymer segments on the cis side of the membrane. With these insights, results from Figure 5 suggest that due to the external electric field, polymers are elongated at the onset of translocation yet compress down to a size comparable to their corresponding full-length relaxed conformations by the time the tension front is furthest from the pore (Figure 5d). Section S7 of the SI further expands on the role of polymer conformation on velocity profiles by quantifying the correlation between inter-segment velocities and its dependence on contour location.

The translocation velocity profiles experimentally obtained in this work as a function of pore size and polymer length allowed us to elucidate the forces at play inside and outside the nanoscale confinement of a nanopore during the polymer translocation process. The use of the nanostructured velocity profiling molecules (VMPs) removed the need for complex instrumentation and avoided neglecting the time-dependent forces resulting from the non-equilibrium nature of strongly driven translocations which were limitations of previous studies.[17,24] The proposed physical toy model allowed us to convert velocity fluctuations into force measurements, and experimentally validate theoretical concepts from Tension Propagation regarding the dependence of velocity profiles on polymer length and pore sizes which, despite being hinted at by simulations, had eluded thorough experimental validation for a decade. Section S8 of the SI deepens this analysis by reporting the dependence of profiles on the applied voltage $\Delta V$. Our experiments also displayed the importance of considering external field effects on polymers prior to translocation, as often omitted by theoretical models. Knowledge of the dependence of the velocity profiles on experimental parameters (pore size, applied voltage) can be used to optimize operating condition for a range of biosensing applications, including DNA nanostructures as barcodes for multiplexed biomarker detection or for digital information storage. The VPM-based method introduced can be easily extended to study the nanoscale forces present in different nanopore sensing experimental conditions and device architectures.

## Methods

### *Linearized Scaffold Preparation*

M13mp18 circular single-stranded DNA (New England Biolabs, N4040S) were linearized to create a 7249 nt long single-stranded DNA. First, primers were attached to the circular scaffold by mixing them with 10 $\mu g$ of M13mp18 at a ratio of 10:1 in 1x NEB 3.1 buffer, heating the sample to 95 $°C$, and slowly cooling down to room temperature in a MiniAmp Plus Thermal Cycler (ThermoFisher Scientific, #A37835). To linearize the scaffolds, a mixture of circular scaffold (with primer attached), and 10 units of HincII restriction enzyme (New England Biolabs, R0103S) in a total reaction volume of 50 $\mu L$ in 1x NEB 3.1 buffer were incubated at 37 $°C$ for 3 hours, then heat-inactivated at 65 $°C$ for 20 minutes in the thermal cycler.

*Nanostructure Assembly*

For the assembly of the VPMs, the linearized scaffold was mixed with 171 staple strands at a molar ratio of 1:10 in assembly buffer (40 mM Tris, 20 mM acetic acid, 2 mM EDTA, and 16 mM $MgCl_2$, pH 8). The product was heated to 95 $°C$ for 5 minutes, then cooled to 90 $°C$, ramped from 90 $°C$ to 60 $°C$ at a rate of 0.4 $°C$ per minute, then from 60 $°C$ to 26 $°C$ at a rate of 0.03 $°C$, and snap cooled to 4 $°C$ using a minicamp Plus Thermal Cycler. After assembly, nanostructures purified using 100 kDa Amicon Ultra-0.5 Centrifugal Filter Unit (Millipore Sigma, UFC500396). Three washes with the assembly buffer were performed to completely remove excess staple strands present in the solution. The assembled products were visualized on 0.5% agarose gel in 1x TAE buffer, as shown in section S1 of the SI.

*Nanopore Fabrication*

Nanopores were fabricated in 12 nm thick free-standing $SiN_x$ membranes (Norcada, NBPX5004Z) using the controlled breakdown (CBD) method. Pores were fabricated in 1 M KCl at pH 8 and slowly grown to desired sizes with AC voltages pulses of small amplitudes in 3.6 M LiCl, as described by Waugh *et al.*[43] Prior to fabrication, membranes were painted with a layer of PDMS to reduce high-frequency noise.[44]

*Nanopore Sensing*

A volume of 1-2 $\mu L$ of the VPM samples were added to $50 - 80\ \mu L$ of 3.6 M LiCl solution with pH 8 for nanopore sensing. Prior to all VPM experiments, 5 kbp dsDNA (ThermoFisher Scientific, SM1731) was run as a control experiment. The ionic current recordings were either performed using the VC100 current amplifier (Chimera Instruments) with a bandwidth of 1 MHz for nanopores with baseline currents smaller than 20 nA, or using the Axopatch 200B amplifier with a bandwidth of 100 kHz for larger pores.

*Data Analysis*

Translocation events in the recorded current traces were located and fitted using a custom implementation of the CUSUM+ algorithm.[45] A digital low-pass filter was applied prior to event detection and fitting. The value of the cutoff frequency was chosen on a per-experiment basis, so as to maximize temporal resolution whilst maintaining good signal to noise ratio. The fitted translocation events were plotted and further analyzed using Nanolyzer (v0.1.41) from Northern Nanopore Instruments

**Acknowledgements**

M. C. acknowledges support of the Ontario Graduate Scholarship (OGS). All authors would like to acknowledge the support of the Natural Sciences and Engineering Research Council of Canada (NSERC), [funding reference number RGPIN-2021-04304], and thank Dr. Kyle Briggs for the insightful discussions leading up to this work.

# Supporting Information: Velocity fluctuation and force scaling during driven polymer transport through a nanopore

Martin Charron, Breeana Elliott, Nada Kerrouri, Liqun He, Vincent Tabard-Cossa*

150 Louis-Pasteur Private, Department of Physics, University of Ottawa, Ottawa K1N 6N5, Canada

*Corresponding Author: tcossalab@uottawa.ca

## S1. DNA structure design and sequences

### Scaffold preparation and sequence

The linearized M13 single-stranded DNA scaffolds were prepared as described in the Methods Section in the main text, using M13mp18 circular single-stranded DNA (New England Biolabs, N4040S). A primer strand was added in a mixture with 10 $\mu g$ M13mp18 circular single-stranded DNA at a ratio of 10:1 in 1x NEB 3.1 buffer, the mixture was heated to 95 $^{\circ}C$ and slowly cooled down to room temperature in a thermal cycler. A mixture of prepared circular scaffold (with primer attached), and 10 units of HincII restriction enzyme (New England Biolabs, R0103S) in a total reaction volume of 50 $\mu L$ in 1x NEB 3.1 buffer were incubated at 37 $^{\circ}C$ for 3 hours, then heat inactivated at 65 $^{\circ}C$ for 20 minutes in a thermal cycler. The sequence of the M13mp18 scaffold is shown below, showing linearization by HincII (New England Biolabs, R0103S). The underlined sequence represents the region where the primer strand is attached, and the red sequences are the recognition site for HincII restriction enzyme.

GACCTGCAGGCATGCAAGCTTGGCACTGGCCGTCGTTTTACAACGTCGTGACTGGGAAAACCCTGGCGTTACCCAACTTAATCGCCTTG
CAGCACATCCCCCTTTCGCCAGCTGGCGTAATAGCGAAGAGGCCCGCACCGATCGCCCTTCCCAACAGTTGCGCAGCCTGAATGGCG
AATGGCGCTTTGCCTGGTTTCCGGCACCAGAAGCGGTGCCGGAAAGCTGGCTGGAGTGCGATCTTCCTGAGGCCGATACTGTCGTCGTC
CCCTCAAACTGGCAGATGCACGGTTACGATGCGCCCATCTACACCAACGTGACCTATCCCATTACGGTCAATCCGCCGTTTGTTCCCACG
GAGAATCCGACGGGTTGTTACTCGCTCACATTTAATGTTGATGAAAGCTGGCTACAGGAAGGCCAGACGCGAATTATTTTTGATGGCGTTCC
TATTGGTTAAAAAATGAGCTGATTTAACAAAAATTTAATGCGAATTTTAACAAAATATTAACGTTTACAATTTAAATATTTGCTTATACAATCTTCCTG
TTTTTGGGGCTTTTCTGATTATCAACCGGGGTACATATGATTGACATGCTAGTTTTACGATTACCGTTCATCGATTCTCTTGTTTGCTCCAGACTC
TCAGGCAATGACCTGATAGCCTTTGTAGATCTCTCAAAAATAGCTACCCTCTCCGGCATTAATTTATCAGCTAGAACGGTTGAATATCATATTGA
TGGTGATTTGACTGTCTCCGGCCTTTCTCACCCTTTTGAATCTTTACCTACACATTACTCAGGCATTGCATTTAAAATATATGAGGGTTCTAAAAA
TTTTTATCCTTGCGTTGAAATAAAGGCTTCTCCCGCAAAAGTATTACAGGGTCATAATGTTTTTGGTACAACCGATTTAGCTTTATGCTCTGAGG
CTTTATTGCTTAATTTTGCTAATTCTTTGCCTTGCCTGTATGATTTATTGGATGTTAATGCTACTACTATTAGTAGAATTGATGCCACCTTTTCAGCT
CGCGCCCCAAATGAAAATATAGCTAAACAGGTTATTGACCATTTGCGAAATGTATCTAATGGTCAAACTAAATCTACTCGTTCGCAGAATTGG
GAATCAACTGTTATATGGAATGAAACTTCCAGACACCGTACTTTAGTTGCATATTTAAAACATGTTGAGCTACAGCATTATATTCAGCAATTAAGC
TCTAAGCCATCCGCAAAAATGACCTCTTATCAAAAGGAGCAATTAAAGGTACTCTCTAATCCTGACCTGTTGGAGTTTGCTTCCGGTCTGGTT
CGCTTTGAAGCTCGAATTAAAACGCGATATTTGAAGTCTTTCGGGCTTCCTCTTAATCTTTTTGATGCAATCCGCTTTGCTTCTGACTATAATAGT
CAGGGTAAAGACCTGATTTTTGATTTATGGTCATTCTCGTTTTCTGAACTGTTTAAAGCATTTGAGGGGGATTCAATGAATATTTATGACGATTCC
GCAGTATTGGACGCTATCCAGTCTAAACATTTTACTATTACCCCCTCTGGCAAAACTTCTTTTGCAAAAGCCTCTCGCTATTTTGGTTTTTATCG
TCGTCTGGTAAACGAGGGTTATGATAGTGTTGCTCTTACTATGCCTCGTAATTCCTTTTGGCGTTATGTATCTGCATTAGTTGAATGTGGTATTCC
TAAATCTCAACTGATGAATCTTTCTACCTGTAATAATGTTGTTCCGTTAGTTCGTTTTATTAACGTAGATTTTTCTTCCCAACGTCCTGACTGGTATA
ATGAGCCAGTTCTTAAAATCGCATAAGGTAATTCACAATGATTAAAGTTGAAATTAAACCATCTCAAGCCCAATTTACTACTCGTTCTGGTGTTT
CTCGTCAGGGCAAGCCTTATTCACTGAATGAGCAGCTTTGTTACGTTGATTTGGGTAATGAATATCCGGTTCTTGTCAAGATTACTCTTGATGAA
GGTCAGCCAGCCTATGCGCCTGGTCTGTACACCGTTCATCTGTCCTCTTTCAAAGTTGGTCAGTTCGGTTCCCTTATGATTGACCGTCTGCG
CCTCGTTCCGGCTAAGTAACATGGAGCAGGTCGCGGATTTCGACACAATTTATCAGGCGATGATACAAATCTCCGTTGTACTTTGTTTCGCG
CTTGGTATAATCGCTGGGGGTCAAAGATGAGTGTTTTAGTGTATTCTTTTGCCTCTTTCGTTTTAGGTTGGTGCCTTCGTAGTGGCATTACGTATT
TTACCCGTTTAATGGAAACTTCCTCATGAAAAAGTCTTTAGTCCTCAAAGCCTCTGTAGCCGTTGCTACCCTCGTTCCGATGCTGTCTTTCGC
TGCTGAGGGTGACGATCCCGCAAAAGCGGCCTTTAACTCCCTGCAAGCCTCAGCGACCGAATATATCGGTTATGCGTGGGCGATGGTTGT
TGTCATTGTCGGCGCAACTATCGGTATCAAGCTGTTTAAGAAATTCACCTCGAAAGCAAGCTGATAAACCGATACAATTAAAGGCTCCTTTTG

GAGCCTTTTTTTTGGAGATTTTCAACGTGAAAAAATTATTATTCGCAATTCCTTTAGTTGTTCCTTTCTATTCTCACTCCGCTGAAACTGTTGAAAG
TTGTTTAGCAAAATCCCATACAGAAAATTCATTTACTAACGTCTGGAAAGACGACAAAACTTTAGATCGTTACGCTAACTATGAGGGCTGTCTG
TGGAATGCTACAGGCGTTGTAGTTTGTACTGGTGACGAAACTCAGTGTTACGGTACATGGGTTCCTATTGGGCTTGCTATCCCTGAAAATGAG
GGTGGTGGCTCTGAGGGTGGCGGTTCTGAGGGTGGCGGTTCTGAGGGTGGCGGTACTAAACCTCCTGAGTACGGTGATACACCTATTCC
GGGCTATACTTATATCAACCCTCTCGACGGCACTTATCCGCCTGGTACTGAGCAAAACCCCGCTAATCCTAATCCTTCTCTTGAGGAGTCTC
AGCCTCTTAATACTTTCATGTTTCAGAATAATAGGTTCCGAAATAGGCAGGGGGCATTAACTGTTTATACGGGCACTGTTACTCAAGGCACTG
ACCCCGTTAAAACTTATTACCAGTACACTCCTGTATCATCAAAAGCCATGTATGACGCTTACTGGAACGGTAAATTCAGAGACTGCGCTTTCC
ATTCTGGCTTTAATGAGGATTTATTTGTTTGTGAATATCAAGGCCAATCGTCTGACCTGCCTCAACCTCCTGTCAATGCTGGCGGCGGCTCTG
GTGGTGGTTCTGGTGGCGGCTCTGAGGGTGGTGGCTCTGAGGGTGGCGGTTCTGAGGGTGGCGGCTCTGAGGGAGGCGGTTCCGGTG
GTGGCTCTGGTTCCGGTGATTTTGATTATGAAAAGATGGCAAACGCTAATAAGGGGGCTATGACCGAAAATGCCGATGAAAACGCGCTACA
GTCTGACGCTAAAGGCAAACTTGATTCTGTCGCTACTGATTACGGTGCTGCTATCGATGGTTTCATTGGTGACGTTTCCGGCCTTGCTAATGG
TAATGGTGCTACTGGTGATTTTGCTGGCTCTAATTCCCAAATGGCTCAAGTCGGTGACGGTGATAATTCACCTTTAATGAATAATTTCCGTCAAT
ATTTACCTTCCCTCCCTCAATCGGTTGAATGTCGCCCTTTTGTCTTTGGCGCTGGTAAACCATATGAATTTTCTATTGATTGTGACAAAATAAAC
TTATTCCGTGGTGTCTTTGCGTTTCTTTTATATGTTGCCACCTTTATGTATGTATTTTCTACGTTTGCTAACATACTGCGTAATAAGGAGTCTTAATC
ATGCCAGTTCTTTTGGGTATTCCGTTATTATTGCGTTTCCTCGGTTTCCTTCTGGTAACTTTGTTCGGCTATCTGCTTACTTTTCTTAAAAAGGGC
TTCGGTAAGATAGCTATTGCTATTTCATTGTTTCTTGCTCTTATTATTGGGCTTAACTCAATTCTTGTGGGTTATCTCTCTGATATTAGCGCTCAATT
ACCCTCTGACTTTGTTCAGGGTGTTCAGTTAATTCTCCCGTCTAATGCGCTTCCCTGTTTTTATGTTATTCTCTCTGTAAAGGCTGCTATTTTCATT
TTTGACGTTAAACAAAAAATCGTTTCTTATTTGGATTGGGATAAATAATATGGCTGTTTATTTTGTAACTGGCAAATTAGGCTCTGGAAAGACGCT
CGTTAGCGTTGGTAAGATTCAGGATAAAATTGTAGCTGGGTGCAAAATAGCAACTAATCTTGATTTAAGGCTTCAAAACCTCCCGCAAGTCGG
GAGGTTCGCTAAAACGCCTCGCGTTCTTAGAATACCGGATAAGCCTTCTATATCTGATTTGCTTGCTATTGGGCGCGGTAATGATTCCTACGA
TGAAAATAAAAACGGCTTGCTTGTTCTCGATGAGTGCGGTACTTGGTTTAATACCCGTTCTTGGAATGATAAGGAAAGACAGCCGATTATTGAT
TGGTTTCTACATGCTCGTAAATTAGGATGGGATATTATTTTTCTTGTTCAGGACTTATCTATTGTTGATAAACAGGCGCGTTCTGCATTAGCTGAA
CATGTTGTTTATTGTCGTCGTCTGGACAGAATTACTTTACCTTTTGTCGGTACTTTATATTCTCTTATTACTGGCTCGAAAATGCCTCTGCCTAAAT
TACATGTTGGCGTTGTTAAATATGGCGATTCTCAATTAAGCCCTACTGTTGAGCGTTGGCTTTATACTGGTAAGAATTTGTATAACGCATATGATA
CTAAACAGGCTTTTTCTAGTAATTATGATTCCGGTGTTTATTCTTATTTAACGCCTTATTTATCACACGGTCGGTATTTCAAACCATTAAATTTAGGT
CAGAAGATGAAATTAACTAAAATATATTTGAAAAAGTTTTCTCGCGTTCTTTGTCTTGCGATTGGATTTGCATCAGCATTTACATATAGTTATATAAC
CCAACCTAAGCCGGAGGTTAAAAAGGTAGTCTCTCAGACCTATGATTTTGATAAATTCACTATTGACTCTTCTCAGCGTCTTAATCTAAGCTATC
GCTATGTTTTCAAGGATTCTAAGGGAAAATTAATTAATAGCGACGATTTACAGAAGCAAGGTTATTCACTCACATATATTGATTTATGTACTGTTTC
CATTAAAAAAGGTAATTCAAATGAAATTGTTAAATGTAATTAATTTTGTTTTCTTGATGTTTGTTTCATCATCTTCTTTTGCTCAGGTAATTGAAATGAA
TAATTCGCCTCTGCGCGATTTTGTAACTTGGTATTCAAAGCAATCAGGCGAATCCGTTATTGTTTCTCCCGATGTAAAAGGTACTGTTACTGTAT
ATTCATCTGACGTTAAACCTGAAAATCTACGCAATTTCTTTATTTCTGTTTTACGTGCAAATAATTTTGATATGGTAGGTTCTAACCCTTCCATTATT
CAGAAGTATAATCCAAACAATCAGGATTATATTGATGAATTGCCATCATCTGATAATCAGGAATATGATGATAATTCCGCTCCTTCTGGTGGTTTC
TTTGTTCCGCAAAATGATAATGTTACTCAAACTTTTAAAATTAATAACGTTCGGGCAAAGGATTTAATACGAGTTGTCGAATTGTTTGTAAAGTCTA
ATACTTCTAAATCCTCAAATGTATTATCTATTGACGGCTCTAATCTATTAGTTGTTAGTGCTCCTAAAGATATTTTAGATAACCTTCCTCAATTCCTTT
CAACTGTTGATTTGCCAACTGACCAGATATTGATTGAGGGTTTGATATTTGAGGTTCAGCAAGGTGATGCTTTAGATTTTTCATTTGCTGCTGGC
TCTCAGCGTGGCACTGTTGCAGGCGGTGTTAATACTGACCGCCTCACCTCTGTTTTATCTTCTGCTGGTGGTTCGTTCGGTATTTTTAATGGC
GATGTTTTAGGGCTATCAGTTCGCGCATTAAAGACTAATAGCCATTCAAAAATATTGTCTGTGCCACGTATTCTTACGCTTTCAGGTCAGAAGG
GTTCTATCTCTGTTGGCCAGAATGTCCCTTTTATTACTGGTCGTGTGACTGGTGAATCTGCCAATGTAAATAATCCATTTCAGACGATTGAGCGT
CAAAATGTAGGTATTTCCATGAGCGTTTTTCCTGTTGCAATGGCTGGCGGTAATATTGTTCTGGATATTACCAGCAAGGCCGATAGTTTGAGTT
CTTCTACTCAGGCAAGTGATGTTATTACTAATCAAAGAAGTATTGCTACAACGGTTAATTTGCGTGATGGACAGACTCTTTTACTCGGTGGCCT
CACTGATTATAAAAACACTTCTCAGGATTCTGGCGTACCGTTCCTGTCTAAAATCCCTTTAATCGGCCTCCTGTTTAGCTCCCGCTCTGATTCT
AACGAGGAAAGCACGTTATACGTGCTCGTCAAAGCAACCATAGTACGCGCCCTGTAGCGGCGCATTAAGCGCGGCGGGTGTGGTGGTTA
CGCGCAGCGTGACCGCTACACTTGCCAGCGCCCTAGCGCCCGCTCCTTTCGCTTTCTTCCCTTCCTTTCTCGCCACGTTCGCCGGCTTT
CCCCGTCAAGCTCTAAATCGGGGGCTCCCTTTAGGGTTCCGATTTAGTGCTTTACGGCACCTCGACCCCAAAAAACTTGATTTGGGTGATG
GTTCACGTAGTGGGCCATCGCCCTGATAGACGGTTTTTCGCCCTTTGACGTTGGAGTCCACGTTCTTTAATAGTGGACTCTTGTTCCAAACTG
GAACAACACTCAACCCTATCTCGGGCTATTCTTTTGATTTATAAGGGATTTTGCCGATTTCGGAACCACCATCAAACAGGATTTTCGCCTGCT
GGGGCAAACCAGCGTGGACCGCTTGCTGCAACTCTCTCAGGGCCAGGCGGTGAAGGGCAATCAGCTGTTGCCCGTCTCACTGGTGAA
AAGAAAAACCACCCTGGCGCCCAATACGCAAACCGCCTCTCCCCGCGCGTTGGCCGATTCATTAATGCAGCTGGCACGACAGGTTTCC
CGACTGGAAAGCGGGCAGTGAGCGCAACGCAATTAATGTGAGTTAGCTCACTCATTAGGCACCCCAGGCTTTACACTTTATGCTTCCGGC
TCGTATGTTGTGTGGAATTGTGAGCGGATAACAATTTCACACAGGAAACAGCTATGACCATGATTACGAATTCGAGCTCGGTACCCGGGGAT
CCTCTAGAGTC

## Structure design

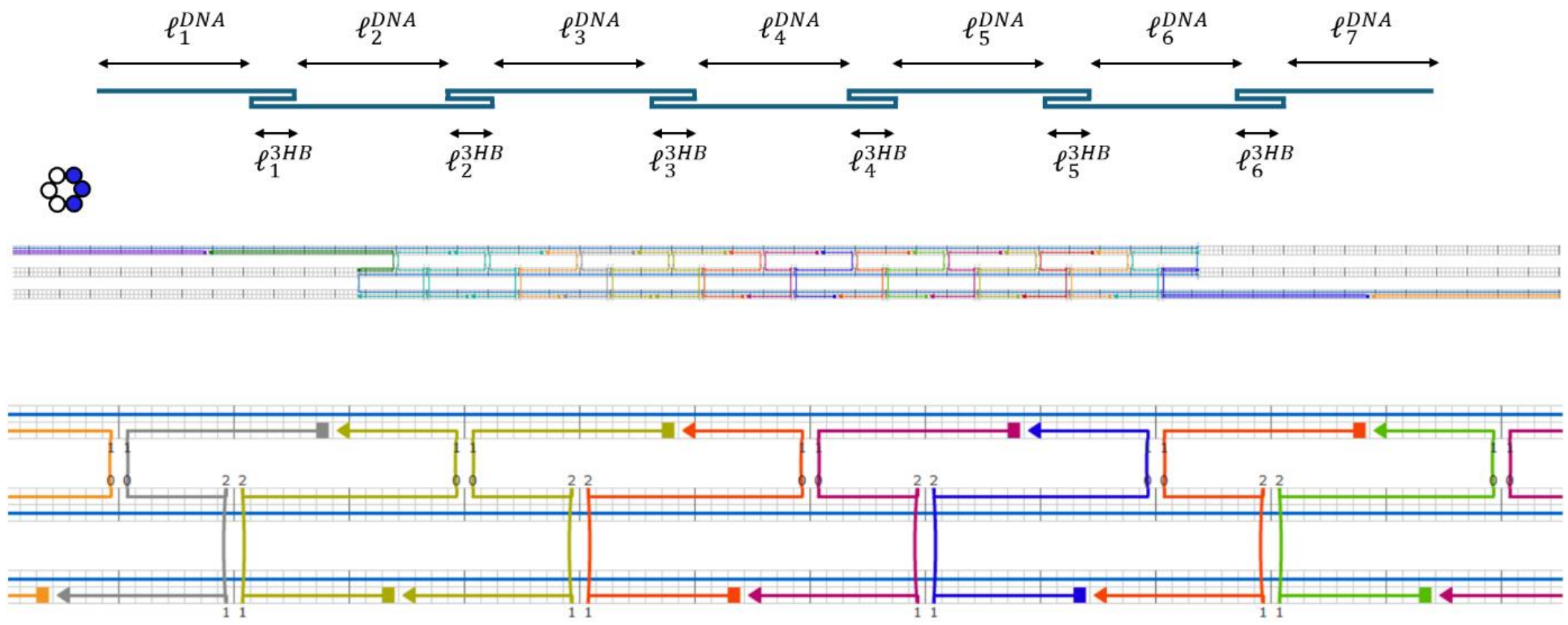

**Figure S1.** caDNAno[1] design of VPM on a honey-comb lattice, using linearized M13mp18 scaffold and 171 staple strands. The long blue strand represents the M13mp18 scaffold, and the short, coloured arrows are the staple strands, with the arrow side being the 3' end.

| Segment | Length (bp) |
|---|---|
| $\ell_1^{DNA}$ | 568 |
| $\ell_1^{3HB}$ | 196 |
| $\ell_2^{DNA}$ | 567 |
| $\ell_2^{3HB}$ | 182 |
| $\ell_3^{DNA}$ | 535 |
| $\ell_3^{3HB}$ | 186 |
| $\ell_4^{DNA}$ | 535 |
| $\ell_4^{3HB}$ | 182 |
| $\ell_5^{DNA}$ | 536 |
| $\ell_5^{3HB}$ | 185 |
| $\ell_6^{DNA}$ | 567 |
| $\ell_6^{3HB}$ | 193 |
| $\ell_7^{DNA}$ | 568 |
| Total | 5000 |

**Table S1.** Lengths of the different segments of the VPM assembly. Segments near extremities were designed slightly longer in anticipation of faster translocation velocities in those areas.

## Nanostructure assembly

| Oligo | Sequence | Length |
|---|---|---|
| 1 | TATAGTCAGAAGCAAAGCGGATTGCATCAAAAAGATTAAGAGGAAGCCCGAAAGAC | 56 |
| 2 | AAAGATTCATCAGTTGAGATTTAGGAATACCACATTCAACTAATGCAGA | 49 |
| 3 | ACATACGAGCCGGAAGCATAAAGTGTAAAGCCTGGGGTGCCTAATGAGT | 49 |
| 4 | AAGGAATTGCGAATAATAATTTTTTCACGTTGAAAATCTCCAAAAAAAAGGCTCCA | 56 |
| 5 | GCTTAATTGAGAATCGCCATATTTAACAACGCCAACATGTAATTTAGGCAGAGGCA | 56 |
| 6 | GGAGCACTAACAACTAATAGATTAGAGCCGTCAATAGATAATACATTTGAGGATTT | 56 |
| 7 | CGTTAATAAAACGCAGACGGTCGAAATCCGCGACCTGCTCCATGTT | 46 |
| 8 | TAATTTCATCTTCTGACCTAAATTTAATGGTTTGAAATACCGACCGTGTGATAAAT | 56 |
| 9 | TCGCTATTAAACAATTTCATTTGAATCCTGAT | 32 |
| 10 | CCGATATATTCGGTCGCTGAGGCTTGCAGGGAGTTAAAGGCCGCTTTTGCGGGATC | 56 |
| 11 | CGGTAATCGTCCTCAGAGCATAAAAATTCTAC | 32 |
| 12 | GGCTTAGGTTCGGATTCGCCTGATTACAGTAACAGTACCTTTTACATCGG | 50 |
| 13 | TGCCAGTTTGAGGGGACGACGACAGTATCGGCCTCAGGAAGATCGCACTCCAGCCA | 56 |
| 14 | ATTTTCGGTCATAGCCCCCTTATTAGCGTTTGCCATCTTTTCATAATCAAAATCAC | 56 |
| 15 | TTATAAATCAAAAGAATAGCCCGAGATAGGGTTGAGTGTTGTTCCAGTTTGGAACA | 56 |
| 16 | CCCTAAAGGGTTTTATAATCAGTGATCACTTG | 32 |
| 17 | AAAATTCATATGGTTTACCAGCGCCAAAGACAAAAGGGCGACATTCAACCGATTGA | 56 |
| 18 | AGCTACAATGCCGAACAAAGTTACCAGAAGGAAACCGAGGAAACG | 45 |
| 19 | AGAGTCCACTATTAAAGAACGTGGACTCCAACGTCAAAGGGCGAAAAACCGTCTAT | 56 |
| 20 | TAACCCTCGTTTACCAGACGACGATAAAAACCAAAATAGCGAGAGGCTT | 49 |
| 21 | CGTCAGATGAATATGCTTTGTTTTAACCTCC | 31 |
| 22 | GCAAACAAGACGGTTGTACCAAAAGCGAGCTG | 32 |
| 23 | GCATTAGACGGGAGAATTAACTGAACAGAGAATCGTTTTTATTT | 44 |
| 24 | AACCACCACAGTACTATGGTTGCTCTGAAATG | 32 |
| 25 | TTGCGGATGGCTTAGAGCTTAATTGCTGAATATAATGCTGTAGCTCAACATGTTTT | 56 |
| 26 | ATTCATCAATATAATTACCTTCTGTAAATCG | 31 |
| 27 | GACGCTCAATCGTTTGACGAACGCTGCGCGT | 31 |
| 28 | TGACCGTAATGGGATAGGTCACGTTGGTGTAGATGGGCGCATCGTAACCGTGCATC | 56 |
| 29 | ACAGCTTGATACCGATAGTTGCGCCGACAATGACAACAACCATCGCCCACGCATAA | 56 |
| 30 | TAATGGAAGGGTTATCAAGATGAAAACATAG | 31 |
| 31 | CAATAATAACGGAATACCCAAAAGAACTGGCATGATTAAGACTCCTTAT | 49 |
| 32 | TGGGCTTGAGATGCTTGACAAGAGGCAAAAG | 31 |
| 33 | TCATCGTAGGAGCAGCCTTTACAGACCCTGAA | 32 |
| 34 | AATACACTAATCATCAAGAGTAATGTTTAATT | 32 |
| 35 | GTAAAACAGAAATTTCATTTCAATAGTGAAT | 31 |
| 36 | TCATGGAAATACCCTTTCCTGCGCTGGCAAG | 31 |
| 37 | ACTTAGCCGGAACGAGGCGAACTAACGGAACAACATTATTACAGGTAG | 48 |
| 38 | TTATCAAAATGCAGAGGCGAATTAAAAGAAAT | 32 |
| 39 | TGTAGCGGTCGCACGTATAACGTGTACATTTT | 32 |
| 40 | CGATAGCTTAATGATGAAACAAACAGAACCTA | 32 |
| 41 | CAACAGGTCAGGATTAGAGAGTACCTTTAATTGCTCCTTTTGATAAGAGGTCATTT | 56 |
| 42 | TTAGAATCCTAAACAAAATTAATTCTTCTGAA | 32 |
| 43 | TATCGGCCTTGCTAACGGTACGGGGAAAGCC | 31 |
| 44 | TAAAATTCGCATTAAATTTTTGTTAAATCAGCTCATTTTTTAACCAATA | 49 |
| 45 | AGAACCACCACCAAGTGCCCGCCCGGAATAG | 31 |
| 46 | TTGCAAAAGAAGTTTTGCCAGAGGGGGTAATAGTAAAATGTTTAGACTGGATAGCG | 56 |
| 47 | CTAAAACGAAAGAACCGGATATTCTAGTAAAT | 32 |
| 48 | AAGGCAAAGAATTGTACCCCGGTTGATAATCAGAAAAGCCCCAAAAAC | 48 |
| 49 | AACAATAGATAAGTCCTGAACAAGAAAAATAATATCCCATCCTAATTTACGAGCAT | 56 |
| 50 | TCCAGACGACGACAATAAACAACATGTTCAGCTAATGCAGAACGCGCCTGTTTATC | 56 |
| 51 | AATATATGTGAGTGAATAACCTTGCTTTTTTAATGGAAACTGATGGCA | 48 |
| 52 | GAGCTAACTCACATTAATTGCGTTGCGCTCACTGCCCGCTTTCCAGTCG | 49 |
| 53 | CCACAAGAATTGACCCAATCGCTTATCCGGT | 31 |
| 54 | AGTCTCTGAATTTACCCCAGAATGGAAAGCGC | 32 |
| 55 | TTGCCCCAGCAGGCGAAAATCCTGTTTGATGGTGGTTCCGAAATCGGCAAAATCCC | 56 |
| 56 | CAAAGTCAGAGGGTGAAAATAATCATTACCG | 31 |
| 57 | ATTCTAAGAAGCCATATTATTTATGTTAAGCC | 32 |
| 58 | GCTTTCCGGCACCGCTTCTGGTGCCGGAAACCAGGCAAAGCGCCATTCGCCATTCA | 56 |
| 59 | AACCGCCACCGATGATACAGGAGTGATTGGCC | 32 |
| 60 | GCCAGCATTGACAGGGGTCAGTACTCAGGAG | 31 |
| 61 | TTCAAATATCGCGTTTTAATTCGAGCTTCAAAGCGAACCAGACCGGAAGCAAACTC | 56 |
| 62 | TGCCACTACGAAATCAACGTAACAGACGAGAA | 32 |
| 63 | CCAGAACAATATTATTAAAGGCGAGAAAGGA | 31 |
| 64 | TGGCCAACAGAGATAGAACCCTTCTGACCTGAAAGCGTAAGAATACGTGGCACAGA | 56 |
| 65 | AAAGGAGCCTTTAATTGTATCGGTTTATCAGCTTGCTTTCGAGGTGAATTTCTTAA | 56 |
| 66 | GATTATTTACATTAGGGCGCCCCGCCGCGCTTAATG | 36 |
| 67 | ACACCAGAACGAGATTACCCAAGGCACCAAC | 31 |
| 68 | TAATAGTAGTAGCAATAAAGAAAACTAGCAT | 31 |
| 69 | CAAGAGAAGGATTAGGAGAGGCTGAGACTCCT | 32 |
| 70 | TACATAACGCCAAAAGGAATTACGAGGCATAGTAAGAGCAACACTATCA | 49 |
| 71 | CGTTATTAATTTTAAAAGTTTGAGTAACATTATCATTTTGCGGAACAAAGAAACCA | 56 |
| 72 | CAGTTGGCAAATCAACAGTTGAAAGGAATTGAGGAAGGTTATCTAAAATATCTTTA | 56 |
| 73 | CGCCGCTACGGCAGATTCACCAGTCACACGACCAGTAATAAAAGGGACATTC | 52 |
| 74 | TTCAGGGATAGCAAGCCCAATAGGAACCCATGTACCGTAACACTGAGTTTCGTCAC | 56 |
| 75 | ATCTAAAGCATCACCTTGCTGAACCTCAAATATCAAACCCTCAATCAATATCTGGT | 56 |
| 76 | CAGTACAAACTACAACGCCTGTAGCATTCCACAGACAGCCCTCATAGTTAGCGTAA | 56 |
| 77 | ATCAATATGATGCCTGAGTAATGTTATAACAG | 32 |
| 78 | GGCGAACGTGGGATTTTAGACAGGGGTAATAT | 32 |
| 79 | CCTGAGTAGAAGAGAAGTGTAGCCCCCGATT | 31 |
| 80 | AAATATGCAACTAAAGTACGGTGTCTGGAAGTTTCATTCCAGTAGGTAGTCAAATCACC | 59 |
| 81 | CAATAATAAGAGCATAAACACGCGAGGCGTT | 31 |
| 82 | TAGAGCTTGACGCCAGAATCCTGAACTCAAAC | 32 |
| 83 | AGGAAGATTGTATAAGCAAATATTTAAATTGTAAACGTTAATATTTTGT | 49 |
| 84 | CTGGCTCATTATAAGAGGACCAAAGTACAAC | 31 |
| 85 | AGGCAGGTCAGACGTACTGGGCCACCCTCAG | 31 |
| 86 | ACCACCCTCAGAGTGCCCCCGAGAGGGTTGA | 31 |
| 87 | TTAGCGAACCTTGCCAGTTACAAAAAGAAACA | 32 |
| 88 | CCATATCAAAATTAAAGAAGGATTAAGACGC | 31 |
| 89 | AGGGAAGAAATAAACAGGAGGCCGACCGCCAG | 32 |
| 90 | GGAGATTTGTTGACCAACTTTGAACCAGTCAG | 32 |
| 91 | TATAAGTATAGTATAAACAGTTAACCGCCACC | 32 |
| 92 | TGAGAAGAGTCAATTACCTGAGCAATTTGCAC | 32 |
| 93 | GATTAGTAATAACAGGCCACCTAAATCGGAA | 31 |
| 94 | CGCTAATATCAGATTTTTTGCAAGCAAATCA | 31 |
| 95 | CCGTAAAGCACGAGTAAAAGAGTCACTTCTTT | 32 |
| 96 | GAGTAGATTTAGTTTAGAACAAATTAATGCC | 31 |
| 97 | ATGAAATAGCAATGCCTAATTCCCGACTTGC | 31 |
| 98 | TTTTCGAGCCAGTAATAAGAGAATATAAAGTACCGACAAAAGGTAAAGTAATTCTG | 56 |
| 99 | GTATCATATGCGTTATACAAATTCTTACCAGTATAAAGCCAACGCTCAACAGTAGG | 56 |
| 100 | AAGGCGTTAAATAAGAATAAACACCGGAATCATAATTACTAGAAAAAGCCTGTTTA | 56 |
| 101 | ACCGACTTGAGCCATTTGGGAATTAGAGCCAGCAAAATCACCAGTAGCACCATTAC | 56 |
| 102 | ATATAAAAGAAACGCAAAGACACCACGGAATAAGTTTATTTTGTCACAATCAATAG | 56 |
| 103 | GGAAACCTGTCGTGCCAGCTGCATTAATGAATCGGCCAACGCGCGGGGAGAGGCGG | 56 |
| 104 | GACGTTGGGAAGAACCGAACATCATCGCCTG | 31 |
| 105 | GGAACGCCATCAAAAATAATTCGCGTCTGGCCTTCCTGTAGCCAGCTTTCATCAAC | 56 |
| 106 | ATTAAATGTGAGCGAGTAACAACCCGTCGGATTCTCCGTGGGAACAAACGGCGGAT | 56 |
| 107 | TGTTTGGATTATAACATTTATTAATTTTCCC | 31 |
| 108 | ATAAATTGTGTCAATCATAAGGGAAAAATCTA | 32 |
| 109 | AAGTAAGCAGATATTTATCCTTGCTATTTTGCACCC | 36 |
| 110 | CGCCATTAAAAATACCGAACGAACCACCAGCAGAAGATAAAACAGAGGTGAGGCGG | 56 |
| 111 | GACTCTAGAGGATCCCCGGGTACCGAGCTCGAATTCGTAATCATGG | 46 |
| 112 | TAACCGTTGTAGCAATTGTCCATGGGTCGAGGTG | 34 |
| 113 | CCAGAAGGAGCGGAATTATCATCATATTCCTGATTATCAGAAGTACATAAATC | 53 |
| 114 | GGGAGGGAAGGTAAATATTGACGGAAATTATTCATTAAAGGTGAATTATCACCGTC | 56 |
| 115 | GGCTGCGCAACTGTTGGGAAGGGCGATCGGTGCGGGCCTCTTCGCTATTACGCCAG | 56 |
| 116 | TTGATATTCACAAGGCTTTTCTCAGAACCGC | 31 |
| 117 | TCAAGATTAGTGAATCTTACCAACTTTAAGAA | 32 |
| 118 | CAATATTTTTGAATGGCTATTAGTCTTTAATGCGCGAACTGATAGCCCTAAAACAT | 56 |
| 119 | TACCGAAGCCCTTGCTAACGGAAGCCTTAAA | 31 |
| 120 | GTGTATCACCGTGCCTTGAGTAACGAGCCGCC | 32 |
| 121 | TCAGTATTAACACCGCCTGCAACAGTGCCACGCTGAGAGCCAGCAGCAAATGAAAA | 56 |
| 122 | TACCTTATGCGATCAGACCAGCGATTATACC | 31 |
| 123 | TCAACTTTAATCACTGACCTAACACTCATCT | 31 |
| 124 | CCATTGCAACAGGCGGGAGCGCGAAAGGAGC | 31 |
| 125 | CAATCGCAAGACAAAGAACGCGAGAAAACTTTTTCAAATATATTTTAGT | 49 |
| 126 | TTGACCCCCAGGCGCATAGGCTGGTTGTGAAT | 32 |
| 127 | GGGCGCTAGGCGTTAGAATCAGAGAAAAACGC | 32 |
| 128 | GTCACCCTCAGCAGCGAAAGACAGCATCGGAACGAGGGTAGCAACGGCTACAGAGG | 56 |
| 129 | GAGAAACAATAAGGGTTATATAACTATATGTAAATGCTGATGCAAATC | 48 |
| 130 | CACCCTCAGATAAGCGTCATACATACAAATAA | 32 |
| 131 | TCATAGCTGTTTCCTGTGTGAAATTGTTATCCGCTCACAATTCCACACA | 49 |
| 132 | CAATAACCTGTTTTTTGCGGAAGGCTATCAG | 31 |
| 133 | GATATAGAAGCAAATAAGAAACGAGAGATAAC | 32 |
| 134 | CTGGCGAAAGGGGGATGTGCTGCAAGGCGATTAAGTTGGGTAACGCCAGGGTTTTC | 56 |
| 135 | AAGCGCGAAAAGATGAACGGTGTATTTAAGAA | 32 |
| 136 | GTCATTGCCTGACCCTGTAATACTAGCTATAT | 32 |
| 137 | GGAGAGGGTAAAGGATAAAAATTTTTGACCAT | 32 |
| 138 | GGGAGGTTTTAGCGTCTTTCCAGAAGCTATCT | 32 |
| 139 | CCAGTCACGACGTTGTAAAACGACGGCCAGTGCCAAGCTTGCATGCCTGCAGGTC | 55 |
| 140 | TTTCATTTGGGGCACATTATGAGAGTCTGGA | 31 |
| 141 | TGCGTAGATTTTCAAATCGCCATAGGTCTGA | 31 |
| 142 | AAAAGGTGGCATCGCTAAATGAATCGATGAA | 31 |
| 143 | GGTTTTGCTCAGTTTCTGAAACATGAACCACCCTC | 35 |
| 144 | TACGCAGTATGTTAGCAAACGTAGAAAATACATACATAAAGGTGGCAAC | 49 |
| 145 | GTAGAAACCAATCAATAATCGGCTGTCTTTCCTTATCATTCCAAGAACGGGTATTA | 56 |
| 146 | CAGGGCGATGGCCCACTACGTGAACCATCACCCAAATCAAGTTTTTTGCACGCAAAT | 57 |
| 147 | CATTAGCAAGGCCGGAAACGTCACCAATGAAACCATCGATAGCAGCACCGTAATCA | 56 |
| 148 | GTAGCGACAGAATCAAGTTTGCCTTTAGCGTCAGACTGTAGCGCGTTTTCATCGGC | 56 |
| 149 | TGATTGCCCTTCACCGCCTGGCCCTGAGAGAGTTGCAGCAAGCGGTCCACGCTGGT | 56 |
| 150 | GTCAATCATATAGCAAAATTAAGCATTAACATCCAATAAATCATACAGGC | 50 |
| 151 | TTTTGCTAAACAACTTTCAACAGTTTCAGCGGAGTGAGAATAGAAAGGAACAACTA | 56 |
| 152 | GAATAAGGCTTGCCCTAAGCTGCAAAATACGTAA | 34 |
| 153 | AAGTGCCGTCTGCCTATTTCGGAATCAGAGCC | 32 |
| 154 | AGAACCGCCACCCCCTATTAACCAGGCGGAT | 31 |
| 155 | CGATCTAAAGTTTTGTCGTCTTTCCAGACGTTAGTAAATGAATTTTCTGTATGGGA | 56 |
| 156 | CTTTGAGGACTAAAGACTTTTTCATGAGGAAGTTTCCATTAAACGGGTTCATTCAGT | 57 |
| 157 | GTTTAGTACCTAATAAGTTTTAACGGAGGTTG | 32 |
| 158 | TCCAATACTGCGGAATCGTCATAAATATTCATTGAATCCCCCTCAAATGCTTTAAA | 56 |
| 159 | CAGTTCAGAAAACGAGAATGACCATAAATCAAAAATCAGGTCTTTACCCTGACTAT | 56 |
| 160 | GAGATCTACAGAGAAGCCTTTATTCAAATGGT | 32 |
| 161 | TCTAGCTGATCCTCATATATTTTACTGCGAAC | 32 |
| 162 | TTGATTCCCAATTAATGCAATATTCAACCGT | 31 |
| 163 | ATCCTCATTAAAGGTTCCAGGCCACCACCCTCATT | 35 |
| 164 | AGAAGTATTAGACTTTACAAACAATTCGACAACTCGTATTAAATCCTTTGCCCGAA | 56 |
| 165 | GAGACTACCTAATACCAAGTTACAAGGTTTAA | 32 |
| 166 | CGGAACCAGAGCCACCACCGGAACCGCCTCCCTCAGAGCCGAGTATTAATTAGCGG | 56 |
| 167 | TTTGCGTATTGGGCGCCAGGGTGGTTTTTCTTTTCACCAGTGAGACGGGCAACAGC | 56 |
| 168 | TAGATACATTTCGTCAACGCGCTATTTTTGA | 31 |
| 169 | CGCCCAATAGTTTAACGTCAAAAATAATTGAG | 32 |
| 170 | AACCAAGTACCGCACTCATCGAGAACAAGCAAGCAACATAAAAACAGGGAAGC | 53 |
| 171 | AGAAAGGCCGGAGACAAAGATTCAAAAGGGTG | 32 |

**Table S2.** Sequences of oligos used for the VPM assembly.

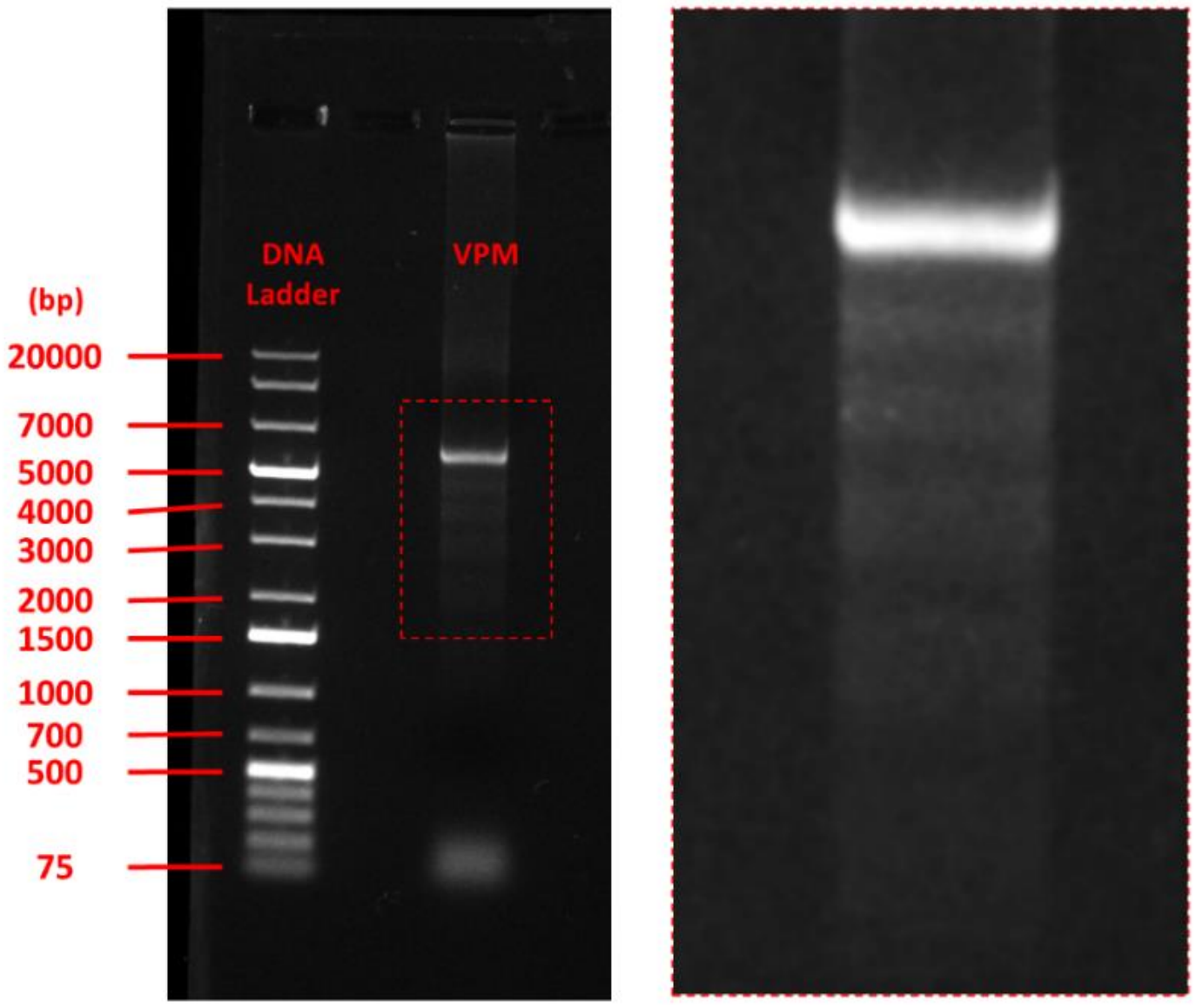

**Figure S2.** Gel Electrophoresis prior to filtering the assembly. Left image shows VPM assemblies and a 1kbp+ DNA ladder. Right image shows a zoomed in image of the VPM lane, showing a quantized streak of shorter assemblies, which are attributed to partial VPM assemblies with a randomly cut scaffold.

Figure S2 shows the migration through a 1% agarose gel of VPMs after the assembly process, and prior to the filtering and washing steps (See Methods section). A clear band can be observed slightly higher than the 5kbp reference, which we assign to fully assembled VPMs. Underneath the VPM band, quantized bands can also be observed ranging from down to 1.5 kbp DNA reference length. We assign these subpopulations to the partial VPM assemblies which, as discussed, correspond to broken scaffolds undergoing the assembly process. Lastly, a band corresponding to excess staple strands can be observed around the 75 bp DNA reference length, which is consistently removed by the three wash steps after assembly.

**S2. Velocity Profile Extraction Robustness**

In this section we aim to show the sensitivity of the threshold-crossing algorithm to analysis parameters and velocity-extraction methods to validate the results shown in the main text and the interpretations drawn from them. This is achieved by comparing different methods of velocity extraction, by varying threshold and hysteresis values, and by reporting the effect of temporal resolution on velocity profiles.

As described and showed in Figure 1 of the main text, the standard analysis method used for all figures in the main text relies on defining a threshold value away from the open-pore baseline and noting the times at which the current crosses the established threshold. When the threshold is set between the blockage states of the 3HB and dsDNA segment blockages, the threshold-crossing times can be used to temporally delimit the passages of the different VPM segments. Knowing the length $\ell_i$ and the measured duration $\tau_i$ of each segment, the mean velocity $\overline{v_i}$ is calculated by fitting the distribution of $\ell_i/\tau_i$ to a normal function, from which the standard deviation $\sigma_v$ can also be extracted.

We first compare two methods of calculating segment velocities. Namely, the standard method described above is compared to the one used by Chen *et al.*,[2] wherein the centers of each 3HB segments are used as reference for temporal and spatial calculations, as depicted in Figure S3a. The time corresponding to 3HB segment center was found by using the middle-point between that segment's threshold crossings. Two translocation velocity profiles are shown in Figure S3a, that of a 9 nm pore, and a 20 nm pore under an applied 200 mV voltage. For larger pores, this technique does not have much of an effect on the absolute velocity values. However, for smaller pores, the 3HB segments interact strongly with the pore walls and as a result the absolute velocity values are significantly smaller for velocities extracted from 3HB-centers.

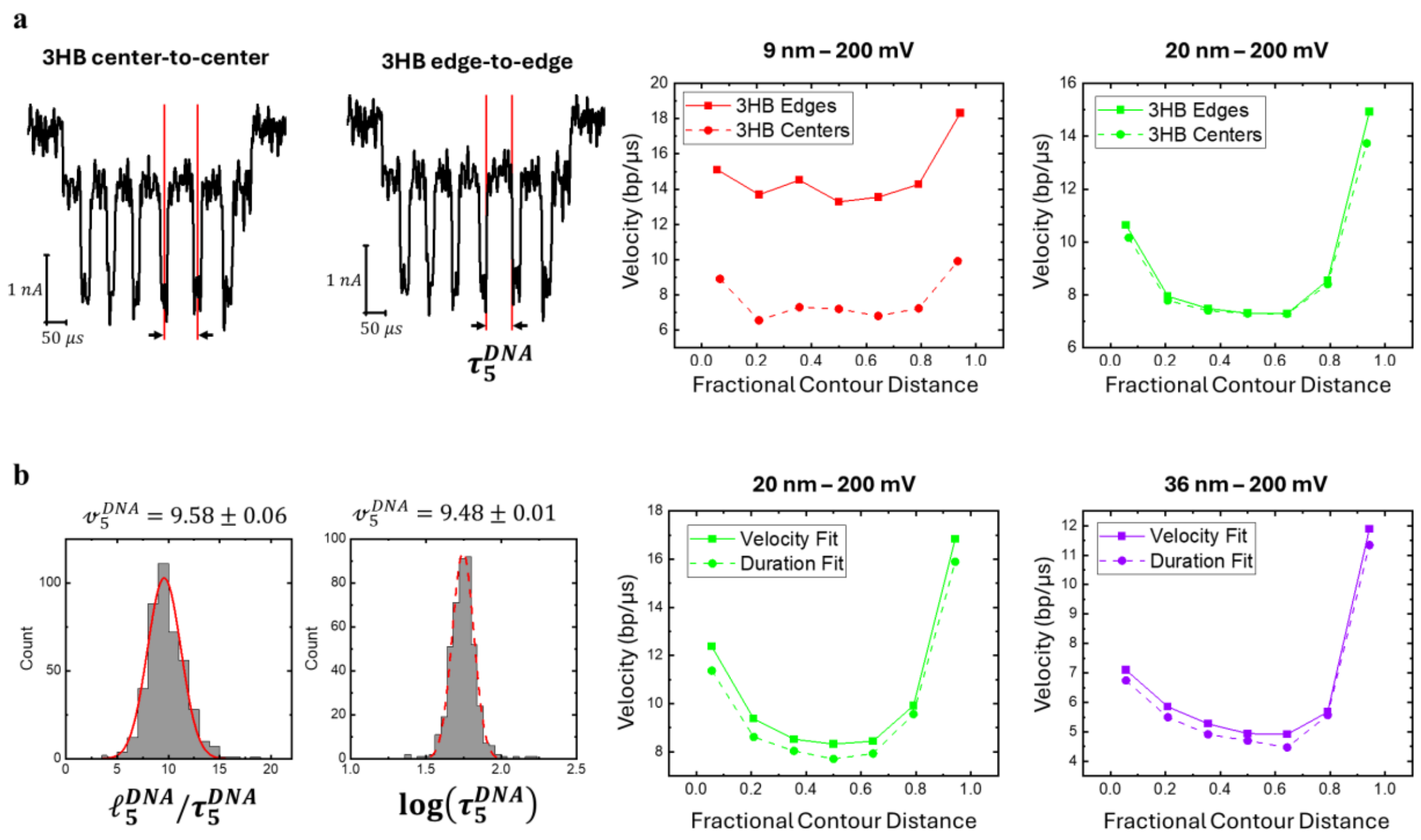

**Figure S3. a)** Sketches displaying the workings of two different duration extraction methods: the 3HB center-to-center and edge-to-edge techniques, as described in the above text. DNA segment velocities are shown for translocations through 9 nm (red) and 20 nm (green) pores. **b)** Comparison of two velocity extraction methods: fitting the distribution of segment velocities calculated using individual segment durations; Fitting the distribution of segment duration to then calculate mean velocity. Comparisons of the two methods are shown for a 20 nm (green) and 36 nm (purple) pore.

Empirically, translocation time distributions are well described by log-normal distributions. Using segment edges as reference, we now compare extracting mean segment velocities $\bar{v}_i$ from normally fitting segment velocities $\langle \ell_i/\tau_i \rangle$ or log-normally fitting segment durations $\tau_i$, extracting the mean segment duration $\bar{\tau}_i$, then calculating the mean velocity as $\bar{v}_i = \ell_i/\bar{\tau}_i$. Figure S3b displays the methods described and plots translocation velocity profiles calculated with both extraction methods for a 20 nm and a 36 nm pore. Velocities from both methods are very similar, although the velocities extracted from duration fits are consistently slightly smaller. Given that both fittings are empirical in nature, we chose to use the velocity fitting method, since unlike log-normal functions, it results in reasonable values for the standard deviations to be used as error bars.

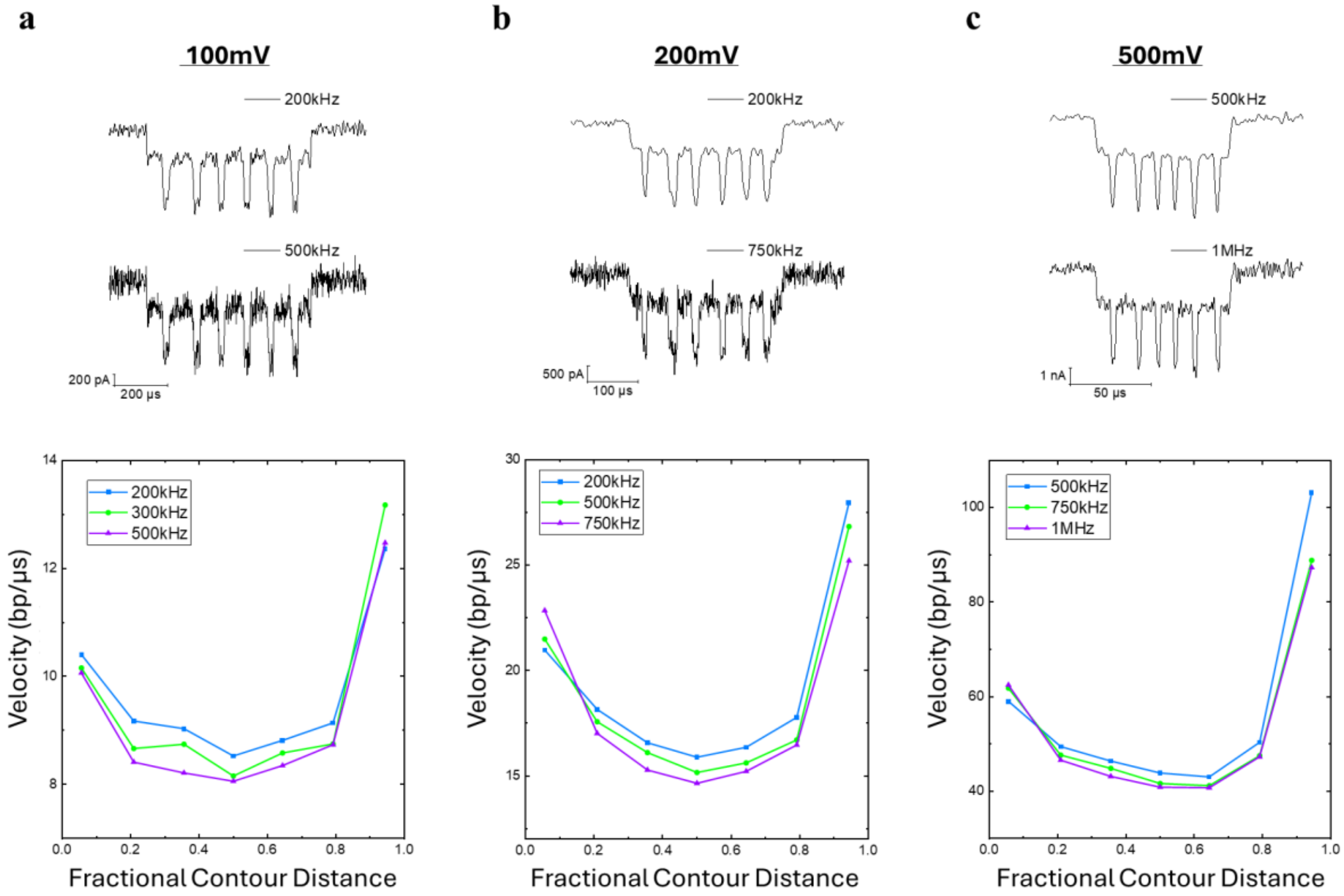

**Figure S4.** Effect of cutoff frequency on velocity extraction for a 11 nm pore. Each figure shows the trace of a VPM event and the extracted velocity profiles for **a)** a voltage of 100 mV, and cutoff frequencies of 200, 300 and 500 kHz, **b)** a voltage of 200 mV, and cutoff frequencies of 200, 500 and 750 kHz, and **c)** a voltage of 500 mV, and cutoff frequencies of 500, 750 and 1,000 kHz.

As the durations of VPM segments get close to the temporal resolution imposed by the signal bandwidth, it's well understood that transient signals become attenuated and deformed. To this end, we now investigate the effects of a signal's temporal resolution on velocity profiles' absolute values and overall shape. As shown in Figure S4, velocity profiles from VPMs passing through an 11 nm pore under voltages of 100, 200 and 500 mV were obtained after filtering current recordings with a Bessel low pass filter with three different cutoff frequency values. Higher cutoff frequencies mostly result in slightly smaller translocation velocities, presumably due to the transitions from DNA-3HB segment transitions being sharper, although values vary within 10% across the different bandwidths explored and the shape of the profiles is not affected significantly.

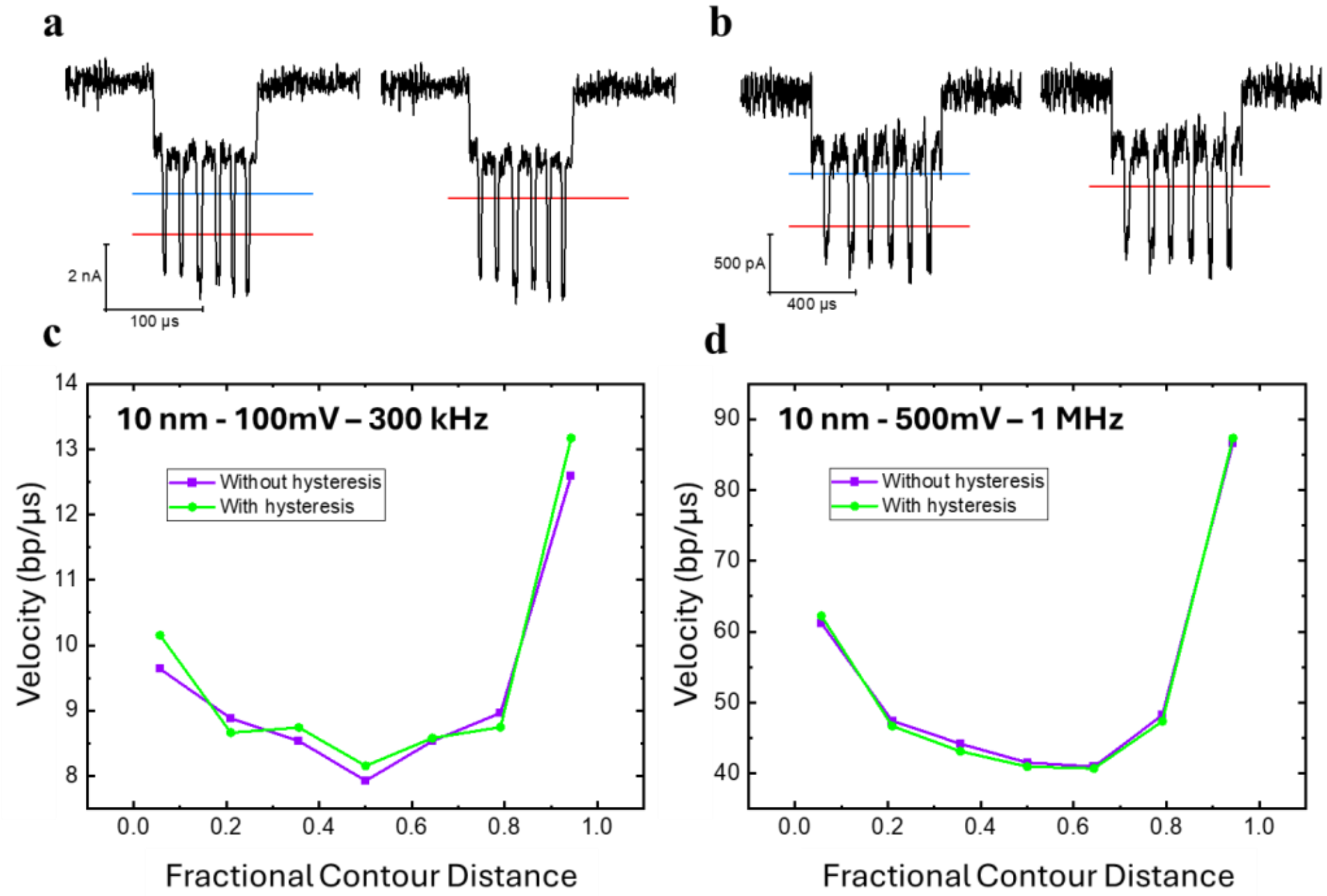

**Figure S5.** Effect of threshold and hysteresis on velocity extraction. **a-b)** Sketch demonstrating the placements of thresholds (red) and hysteresis (blue) for the two different analysis parameters tested. **c-d)** Translocation velocities measured with zero (purple) and non-zero (green) hysteresis for VPM translocations in an 11 nm pore, under a 100 mV voltage (c) and under a 500 mV voltage (d). Cutoff frequencies are denoted in the figures.

We finally test the sensitivity of velocity analysis on threshold parameters, achieved simply by analyzing the same dataset with two different threshold settings. The first is the standard method introduced above, wherein the threshold is used to detect the start of 3HB segments, and the hysteresis is used to detect its end. The second method doesn't use a hysteresis, and as such only the threshold is used to determine the beginning and end of each 3HB, as depicted in Figures S5a-b. Figures S5c-d show the corresponding velocity profiles from translocations through an 11 nm pore under voltages of 100 and 500 mV, respectively, the values of which were extracted with (green) and without (purple) a hysteresis. We note that to analyze events with no hysteresis, a good signal to noise ratio is required as noise from a DNA blockage state could otherwise easily trigger the threshold and be detected as a 3HB segment instead. As such, cutoff filters of 300kHz were used for the analysis of VPM translocations under 100 mV to minimize the events containing false-positive threshold crossings induced by current fluctuations. While increasing temporal resolution would certainly result in more accurate velocity measurements, the 100 mV and 500 mV datasets from Figures S5c and S5d show that the profiles remain mostly unaffected. Note by choosing to evaluate the velocity profiles under two significantly different voltage biases, we tested the limited temporal resolution of higher voltage and the worse signal to noise ratios from lower voltages, thus ensuring the robustness of the velocity measurements to threshold parameters.

Concisely, although not perfect, we conclude from these simple tests that the analysis method used throughout this work constitutes a reliable way to characterize translocation velocity profiles. Namely, if the segment durations are significantly longer than the system's rise time, we suggest velocity profiles do not strongly depend on the analysis parameters.

## S3. Velocity Profiles

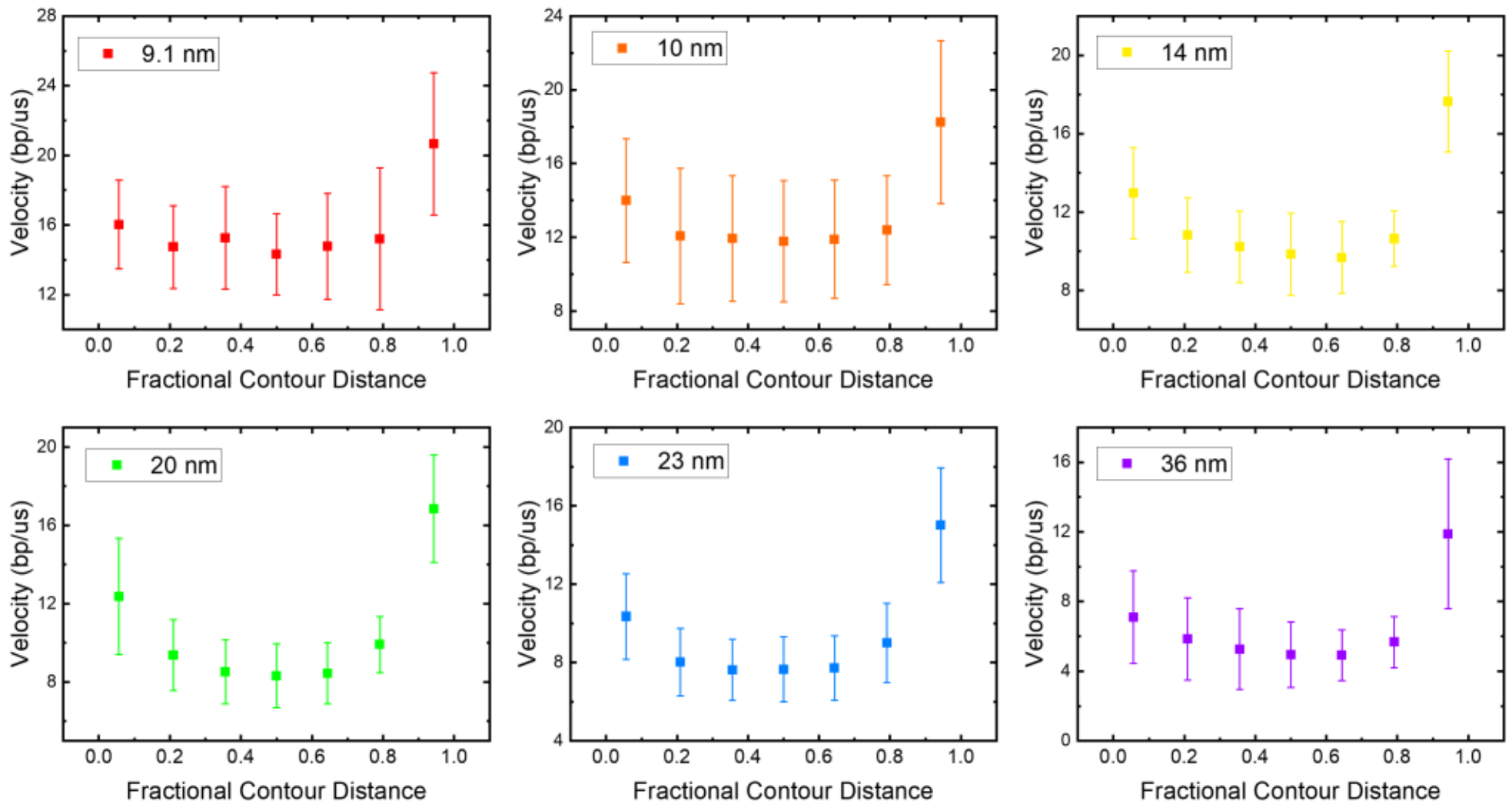

**Figure S6.** Velocity Profiles obtained for different pore sizes, denoted in the legends. Error bars are the standard deviations of the extracted velocities.

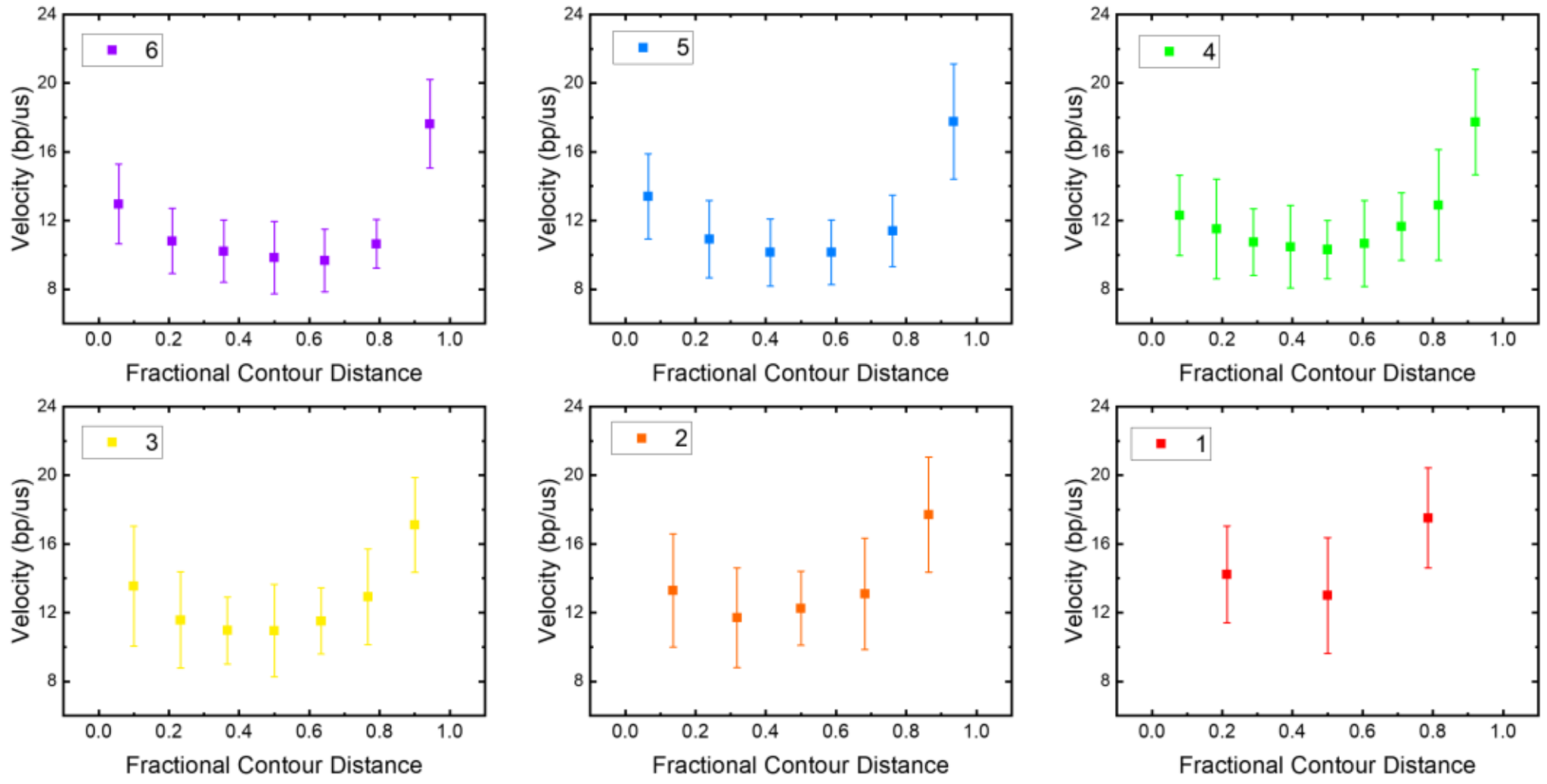

**Figure S7.** Velocity Profiles obtained for different partial VMP assemblies. Legend indicates number of 3HB segments 3HB segment velocities were used for partial VPMS with $n_{3HB} < 5$ to increase spatial resolution. Error bars are the standard deviations of the extracted velocities.

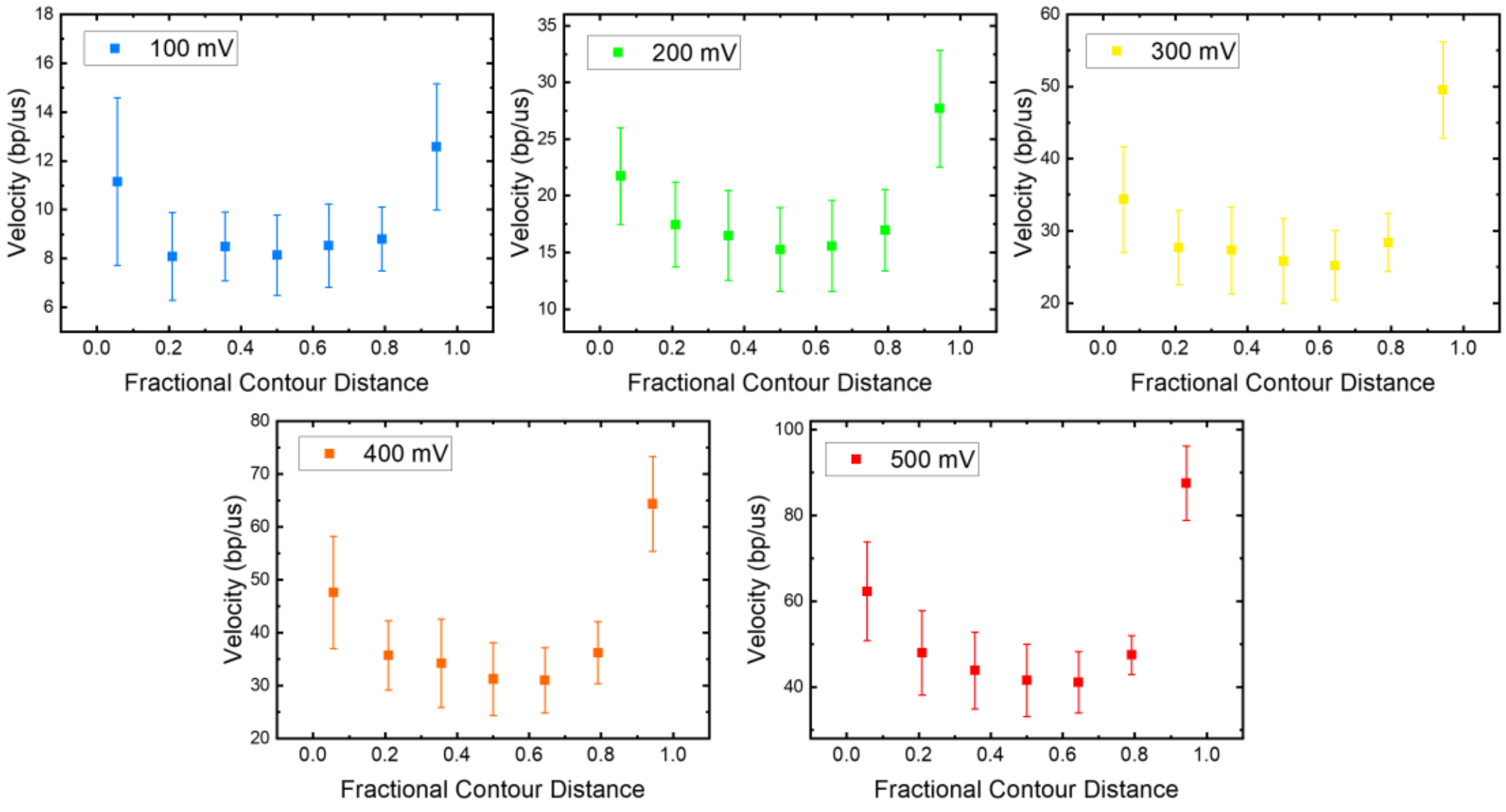

**Figure S8.** Velocity Profiles obtained under different applied voltages. Error bars are the standard deviations of the extracted velocities.

## S4. Translocation Time vs Pore Size

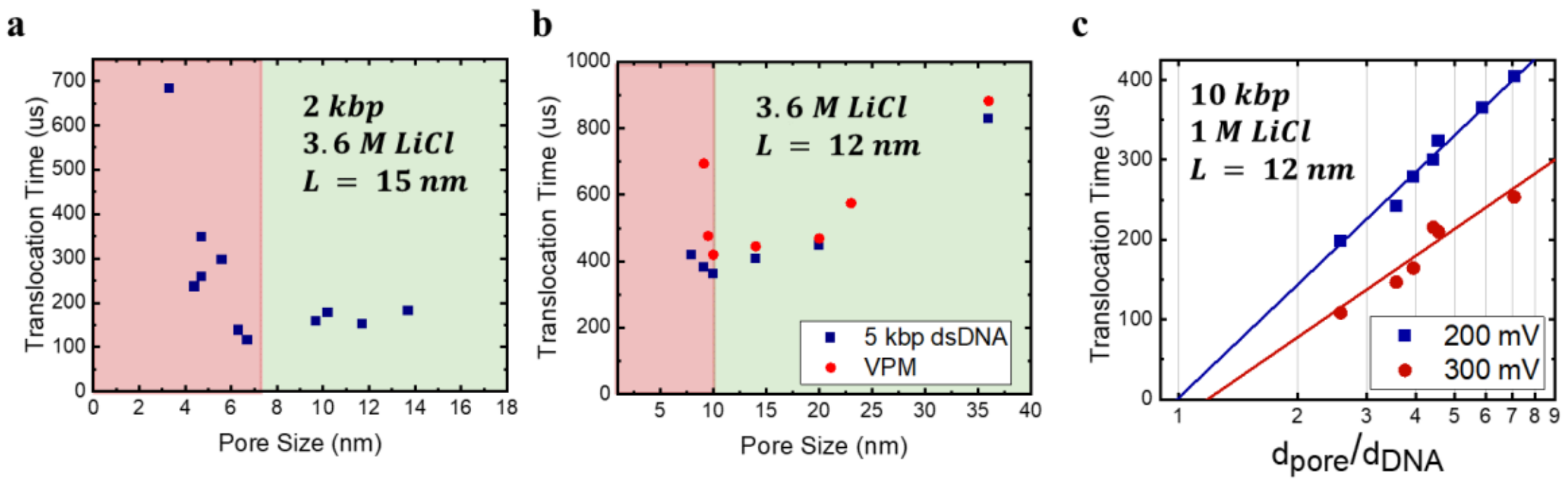

**Figure S9.** Translocation vs pore sizes in different conditions. **a)** 2 kbp dsDNA in 3.6 M LiCl with 15 nm thick membranes. **b)** 5 kbp dsDNA and VPM in 3.6 M LiCl in 12 nm thick membranes. **c)** 10 kbp dsDNA in 1M LiCl with 12 nm thick membranes.

## S5. Expanding on the Toy Model

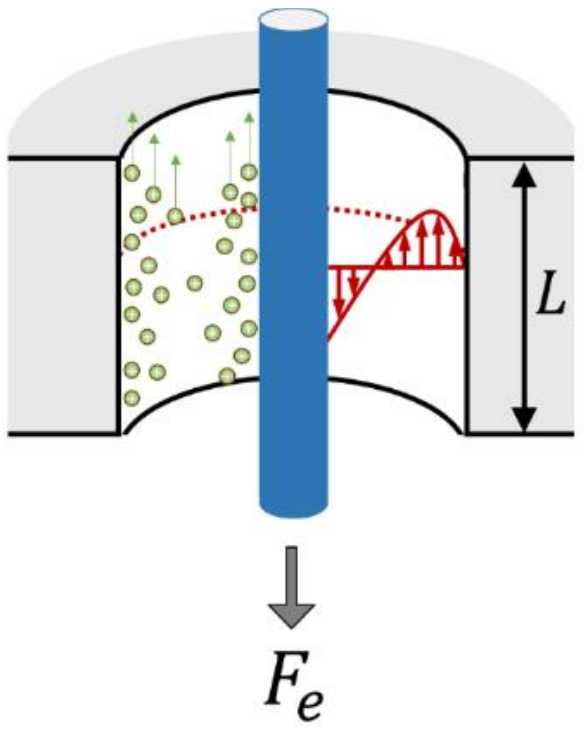

**Figure S10.** Depiction of internal forces involved when pulling a charged cylinder through a charged cylindrical channel.

In the main text, a toy model is used to experimentally characterize the forces in nanopore systems. This was achieved through Equation 3, which required no explicit derivation of individual terms for forces considered in the model. Such derivations can however be achieved for the internal forces under the assumption of working with very long pores, i.e. ignoring end effects and forces outside the pore. The resulting expressions provide useful insights for interpreting translocation velocity results of the main text, as shown here.

As per Figure S10, let a solid cylinder of radius $r_{DNA}$ move through a hollow cylindrical channel of radius $r_{pore}$ at a velocity $\vec{v}_{DNA} \equiv v_{DNA}\hat{z}$ in response to a uniform electrical field $\vec{E} = E\hat{z}$. Furthermore, let the pore walls and moving cylinder have non-zero surface charge densities with corresponding surface potentials of $\phi(r_{pore}) \equiv \phi_{pore}$ and $\phi(r_{DNA}) \equiv \phi_{DNA}$, respectively. Note that cylindrical coordinates are used here, such that $r$ is the radial distance from the pore central axis. The radial dependence of the charge density $\rho(r)$ of counterions between the two charged surfaces is described by the Poisson equation as:

$$\nabla^2\phi(r) = -\frac{\rho(r)}{\varepsilon} \tag{S1}$$

Here, $z$-independence is assumed, and $\varepsilon$ is the bulk's permittivity. Similarly, the radial dependence of the fluid velocity $\vec{v}(r) = v\hat{z}$ between both cylinder's surface is described by the Navier-Stokes equation under an electric field:

$$\eta\nabla^2 v = -\rho(r)E \tag{S2}$$

Here, we consider the no-slip boundary conditions $v(r_{DNA}) = v_{DNA}$ and $v(r_1) = 0$. Combining both differential equations, $\nabla^2 v = \frac{\epsilon E}{\eta}\nabla^2\phi$ , a general expression can be found for the radial fluid velocity profile $v(r)$:

$$v(r) = \frac{\epsilon E}{\eta}\phi(r) + c_1 \ln r + c_2 \tag{S3}$$

Applying the boundary conditions, the final expression for $v(r)$ is found to be:

$$v(r) = \frac{\varepsilon E}{\eta}\left(\phi(r) - \phi_{pore}\right) + \left[v_{DNA} - \frac{\varepsilon E}{\eta}\left(\phi_{DNA} - \phi_{pore}\right)\right]\frac{\ln\frac{r}{r_{pore}}}{\ln\frac{r_{DNA}}{r_{pore}}} \tag{S4}$$

The drag force imparted by the fluid on the moving cylinder of length $L_{DNA}$ can thus be calculated by integrating the viscous stress tensor over moving cylinder's surface, i.e.:

$$F_{drag} = 2\pi r_{DNA} L_{DNA}\eta \left.\frac{dv}{dr}\right|_{r_{DNA}}$$

$$F_{drag} = -Q_{DNA}E + \frac{\varepsilon E}{\eta}\left(\phi_{DNA} - \phi_{pore}\right)\frac{2\pi L_{DNA}\eta}{\ln\frac{r_{pore}}{r_{DNA}}} - \frac{2\pi L_{DNA}\eta}{\ln\frac{r_{pore}}{r_{DNA}}}v_{DNA} \tag{S5}$$

Note that $Q_{DNA}$ corresponds to the total charge of the moving cylinder and arises from Gauss' law, i.e. $Q = \varepsilon 2\pi r_{DNA} L_{DNA}(-d\phi(r_{DNA})/dr)$.

In Equation S5, the first term simply corresponds to the drag force exerted on the charged cylinder under an applied electric field in free solution (i.e. $r_{pore} \to \infty$), perfectly balancing the driving electric force $Q_{DNA}E$ in the opposite direction as expected. Similarly, the first and second term, when combined, correspond to the drag force expected from a stationary charged cylinder ($\mathcal{v}_{DNA} = 0$) stalled inside a charged cylindrical channel, as determined and tested experimentally.[3] Finally, the last term corresponds to the hydrodynamic drag imparted by an uncharged cylinder moving through an uncharged cylindrical channel , i.e. $\phi_{DNA} = \phi_{pore} = 0$. Equation S5 thus shows that drag forces are superimposed and could be rewritten as $F_{drag} = F_{EO} + \gamma_{in}\mathcal{v}_{DNA}$, where

$$\gamma_{in} = \frac{2\pi L_{DNA}\eta}{\ln\frac{r_{pore}}{r_{DNA}}} \propto \left(\ln\frac{d_{pore}}{d_{DNA}}\right)^{-1} \quad \text{(S6)}$$

An expression for $\mathcal{v}_{DNA}$ can be found by balancing the electric pulling force $F_e = Q_{DNA}E$ and the drag forces of Equation 6, $F_e + F_{drag} = 0$:

$$\mathcal{v}_{DNA} = \frac{\varepsilon E}{\eta}\left(\phi_{DNA} - \phi_{pore}\right) = (\mu_{EP} - \mu_{EO})E \quad \text{(S7)}$$

Interestingly, due to the common inverse log dependence of the $(F_e - F_{eo})$ and $\gamma_{in}$ terms, Equation S7 predicts that the steady-state velocity of a charged cylinder moving through a charged cylindrical channel is independent of pore size, unlike the experimental results presented in Figure 3 of the main article and Figure S9. As such, the expected velocity $\mathcal{v}_{DNA}$ simply corresponds to the difference between its bulk electrophoretic velocity $\mu_{EP}E$ and the electroosmotic flow velocity due to the charged pore surface $\mu_{EO}E$.

Because of the superimposed nature of the drag force (Eq. S5), a generic term can be considered for the drag imparted on the segment of length $\ell_{ext}$ under tension outside the pore, i.e. $F_{ext}$, as discussed in the main text:

$$F_{ext} = \gamma_{ext}(\ell)\mathcal{v}_{DNA} = -2\pi C'\eta\ell_{ext}\mathcal{v}_{DNA} \quad \text{(S8)}$$

Here the $C'$ coefficient is an undefined shape factor on the order of unity, and the $2\pi$ factor is pre-emptively used to simplify the derivation of an expression for the steady-state velocity $v_{DNA}$. If we assume that the electric field is completely inside the pore, then we can replace $L_{DNA}$ with the pore length $L_{pore}$ in Equation S5, such that:

$$F_{drag} = -Q_{DNA}E + \frac{\varepsilon E}{\eta}\left(\phi_{DNA} - \phi_{pore}\right)\frac{2\pi L_{pore}\eta}{\ln\frac{r_{pore}}{r_{DNA}}} - \frac{2\pi L_{pore}\eta}{\ln\frac{r_{pore}}{r_{DNA}}}v_{DNA} - 2\pi C'\eta \ell_{ext} v_{DNA} \quad \text{(S9)}$$

The steady state velocity $v_{DNA}(\ell_{ext})$ can thus be found by balancing forces:

$$v_{DNA}(\ell_{ext}) = \frac{\frac{\varepsilon E}{\eta}\left(\phi_{DNA} - \phi_{pore}\right)}{1 + C' \ln\left(\frac{r_{pore}}{r_{DNA}}\right)\frac{\ell_{ext}}{L_{pore}}} \quad \text{(S10)}$$

Unlike Eq. S7, Equation S10 shows that the instantaneous velocity depends on pore size, as expected from experimental results (Fig. 3). Let $s \in [0, L_{DNA}]$ denote the position of the polymer segment inside the pore throughout translocation. Translocation durations can be calculated by integrating $d\tau = ds/\left|v_{DNA}\left(\ell_{ext}(s)\right)\right|$:

$$\begin{aligned}
\tau &= \int_0^{L_{DNA}} \frac{ds}{|v_{DNA}(s)|} \\
&= \frac{1}{\frac{\varepsilon E}{\eta}\left|\phi_{DNA} - \phi_{pore}\right|}\left(L_{DNA} + C' \ln\left(\frac{r_{pore}}{r_{DNA}}\right)\frac{1}{L_{pore}}\int_0^{L_{DNA}} \ell_{ext}(s) ds\right) \\
\tau &= \frac{1}{\frac{\varepsilon E}{\eta}\left|\phi_{DNA} - \phi_{pore}\right|}\left(L_{DNA} + C' \ln\left(\frac{r_{pore}}{r_{DNA}}\right)\frac{\bar{\ell} L_{DNA}}{L_{pore}}\right) \quad \text{(S11)}
\end{aligned}$$

In the final expression of Eq. S11, we introduced $\bar{\ell} \equiv L_{DNA}^{-1}\int_0^{L_{DNA}} \ell_{ext}(s) ds$, the average length of the polymer segment under tension throughout the translocation process. Under the reasonable assumption that $\bar{\ell}$ scales with the polymer radius, $\bar{\ell} \propto R_g \propto L_{DNA}^{\nu}$, we can further write:

$$\tau(L_{DNA}) \approx \frac{1}{\frac{\varepsilon E}{\eta}\left|\phi_{DNA} - \phi_{pore}\right|}\left(L_{DNA} + \frac{C \ln\left(\frac{r_{pore}}{r_{DNA}}\right)}{L_{pore}} L_{DNA}^{1+\nu}\right) \quad \text{(S12)}$$

Equation S12 shares identical dependence on polymer length $L_{DNA}$ as predicted by iso-flux tension propagation principles:[4] $\tau = AL_{DNA} + BL_{DNA}^{1+\nu}$, which shows that the coefficient $\alpha$ measured experimentally ($\tau \sim L_{DNA}^{\alpha}$) depends on whether the $A$ or $B$ coefficient dominates for the experimental conditions used. Eq. S12 shows that longer or narrower pores result in weaker $L_{DNA}^{1+\nu}$ coefficients, thus resulting in smaller $\alpha$ coefficients with values closer to 1. This is expected since the time-independent internal drag forces dominate the process, thus promoting a flatter velocity profile, consistent with arguments presented in the main text.

Additionally, Equations S10 and S12 predict that velocity and translocation times should reduce and increase logarithmically with pore size, respectively, which partly explain the results of Section S4. However, both equations fail to predict why the uniform monotonic reduction with increasing pore size observed in Figure 3. Equation S10 instead predicts that the end velocity should be pore-size independent, while only $\ell_{ext} \neq 0$ measurements from middle segments should be pore-size dependent. These inaccuracies most likely arise from ignoring the end-effects of the channel. For instance, the potential drop does not occur solely inside the pore, but instead also extends into the access regions, an effect which is more important in larger pores. This is also true of the electroosmotic flow, which is also present outside the pore. The first term $\left(\frac{\varepsilon E}{\eta}\left|\phi_{DNA} - \phi_{pore}\right|\right)^{-1}$ of Eq. S12 is thus expected to oversimply the analysis and could potentially provide another source of pore size dependence when calculated properly.

## S6. Simulated Polymer Conformations

Following Tree *et al.*,[5] conformations of Discrete Worm-Like Chains (DWLC) tethered to a pore, i.e. with 10 bp inserted into a pore, were randomly generated in an attempt to predict the minimal velocity location $x_{min}$. For a polymer with a length of N base pairs (bp), this was achieved by generating N steps of length $\ell_s = 1\,bp$. To represent polymer conformations prior to single-file translocations, the first 10 steps were imposed to be directed along the axis of the pore. After the first 10, the rest of the steps were randomly generated and followed the discrete statistics imposed by wormlike chains. For a chain with a persistence length of $\ell_p$, the angle between two consecutive steps is known to follow the following probability distribution:

$$P(\theta) = \frac{\frac{\ell_p}{\ell_s}}{2\sinh\frac{\ell_p}{\ell_s}} e^{\frac{\ell_p}{\ell_s}\cos\theta} \sin\theta \tag{S13}$$

This probability can be integrated and inverted to be sampled from, with the sampling function being:

$$\theta(r) = \cos^{-1}\left(1 + \frac{\ell_s}{\ell_p} ln\left(1 - r\left(1 - e^{-2\frac{\ell_p}{\ell_s}}\right)\right)\right) \tag{S14}$$

Here, the value $r \in U(0,1)$ is numerically sampled from the uniform distribution and corresponds to randomly sampling the cumulative distribution function, $r = \int_0^\theta P(\theta) d\theta$.

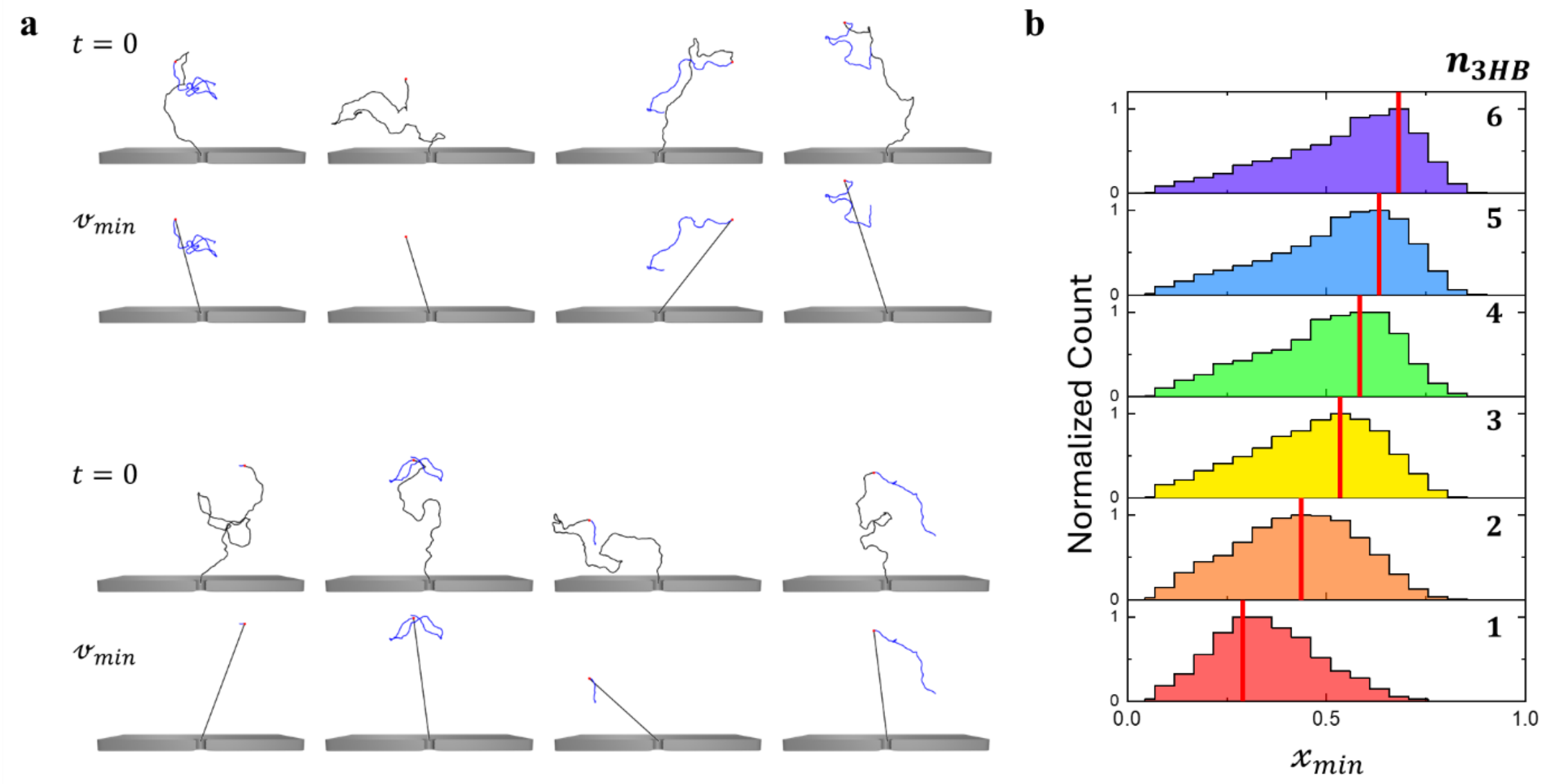

**Figure S11. a)** Simulated polymer conformations shown at two instances: at the onset of translocation, and when the velocity is minimal, i.e. when the tension front reaches the furthest polymer segment. Only segments on the *cis* side are depicted. **b)** Distributions of $x_{min}$ calculated for different polymer lengths.

Figure S11a shows a few of the polymer conformations generated with the above method and with $\ell_p = 150\ bp$. For each generated polymer, the furthest monomer is located, and its distance from the pore center $R_{max}$ is calculated. By then calculating the contour length between the furthest and extremity monomers $\ell_{end}$, the predicted fractional location of the velocity minima was then calculated, as per Equation 4 of the main text, rewritten here:

$$x_{min} = 1 - \frac{R_{max} + \ell_{end}}{L} \quad (S15)$$

Figure S11b shows the distributions of $x_{min}$ values calculated for 5000 simulated polymer conformations. The location of the most probable values from each distribution were noted down, and then used to compare to the experimental velocity profiles in Figure 5 of the main article, showing good agreement between experimental and simulated $x_{min}$ values.

**S7. Segment Duration Correlations**

Following Chen *et al*,[2] we measured the correlation between segment durations within single translocation events. As shown in Figure S12a, this was achieved by calculating the Pearson correlation coefficient $\rho_{ij}$ between $\tau_i^{DNA}$ and $\tau_j^{DNA}$, i.e. the durations of the $i^{th}$ and $j^{th}$ dsDNA segment. Figure S12b displays all the $\rho_{ij}$ values calculated for VPM translocations in a 14 nm pore in a table, where the color intensity of each cell is mapped to the corresponding $\rho_{ij}$ value. As demonstrated by higher values near the table diagonal, the correlation from adjacent segments is strongest than ones further separated. This is expected from neighboring segments reacting to a monomer undergoing an impulse. These interactions are short-range in nature, as the force dissipates throughout the polymer, i.e. a sudden velocity change of one extremity segment won't affect the velocity state of the other extremity of a long polymer.

For better visualization, Figure S12 c shows the correlation values $\rho_{ij}$ plotted against the segment separation $j - i$, where each color corresponds to fixed $j$ value, and varying $i$ values. The correlation between extremity segments, i.e. $j = 1$ and $j = 7$, is consistently weaker than a non-extremity segment with neighboring segments, as observed by the faster decay of $\rho_{i1}$ and $\rho_{i7}$ away from $j - i = 0$ in Figure S12c. This behavior is consistent across the data acquired for this work. We suggest that the higher correlations between non-extremity segments arise from long-range interactions provided by polymer conformations at the onset of translocation, as previously suggested by Lu *et al.*[6] According to Tension Propagation principles, strongly supported by the experiments of this work, a polymer that arrives at the pore in an elongated conformation will have, on average, its monomers farther to the pore than a polymer arriving with a more compressed conformation (Figure S12d). In addition to short range correlations, largely separated segments are expected be correlated through that polymer's conformation at the onset of translocation.

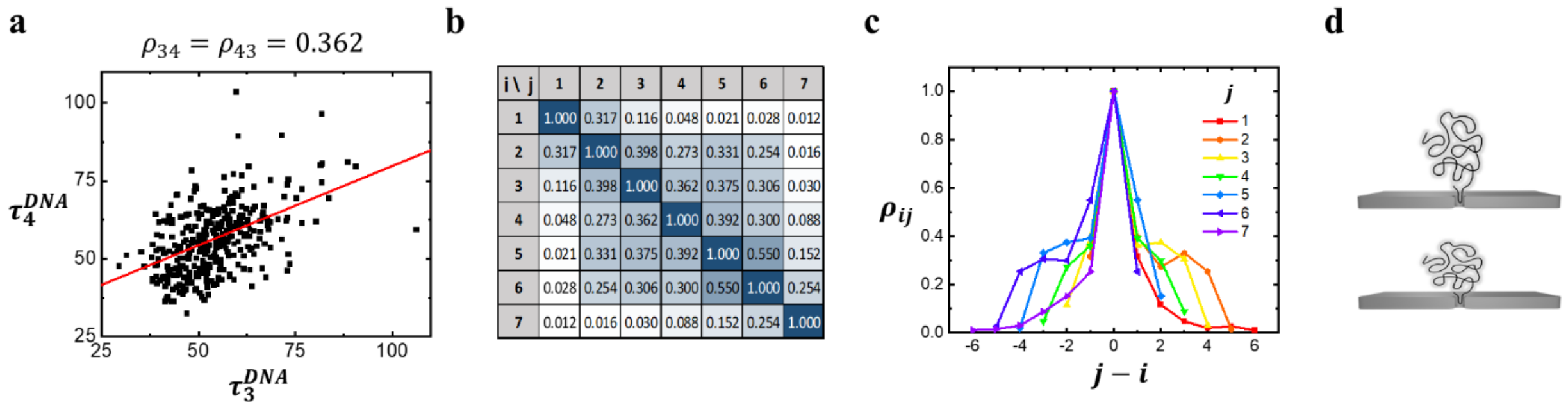

| i\j | 1 | 2 | 3 | 4 | 5 | 6 | 7 |
|---|---|---|---|---|---|---|---|
| 1 | 1.000 | 0.317 | 0.116 | 0.048 | 0.021 | 0.028 | 0.012 |
| 2 | 0.317 | 1.000 | 0.398 | 0.273 | 0.331 | 0.254 | 0.016 |
| 3 | 0.116 | 0.398 | 1.000 | 0.362 | 0.375 | 0.306 | 0.030 |
| 4 | 0.048 | 0.273 | 0.362 | 1.000 | 0.392 | 0.300 | 0.088 |
| 5 | 0.021 | 0.331 | 0.375 | 0.392 | 1.000 | 0.550 | 0.152 |
| 6 | 0.028 | 0.254 | 0.306 | 0.300 | 0.550 | 1.000 | 0.254 |
| 7 | 0.012 | 0.016 | 0.030 | 0.088 | 0.152 | 0.254 | 1.000 |

**Figure S12**. Correlations of segment durations for VPM translocations through a 14 nm pore. **a)** Correlation of $\tau_3^{DNA}$ and $\tau_4^{DNA}$, resulting in the extraction of $\rho_{43} = \rho_{34}$. **b)** Tabulation of $\rho_{ij}$ with cell color intensities mapped to the value of $\rho_{ij}$. **c)** Plot of $\rho_{ij}$ vs $j - i$. Each color corresponds to a fixed value of $j$. **d)** Initial polymer conformations possibly explain long range correlations observed for middle segments undergoing tension propagation.

## S8. Voltage Dependence of Velocity Profiles

Here, we characterize the effect of applied voltage $\Delta V$ on translocation velocity profiles. To this end, VPM passages through a 11 nm nanopore were recorded under voltages of 100, 200, 300, 400 and 500 mV. Traces of each voltage measurements are shown in Figure S13a. Figure S13b shows the distribution of VPM translocation times $\tau$, the mean values of which are plotted in Figure S13c against corresponding $\Delta V$. The mean duration data was fit to a power scaling law of the form $\bar{\tau} \sim \Delta V^{-\beta}$, with a scaling coefficient of $\beta = -1.04 \pm 0.03$ measured. This inverse dependence of $\bar{\tau}$ on voltage agrees with previously published experimental results for dsDNA, which suggests again that VPM translocations should be a good representation of dsDNA translocation kinetics.

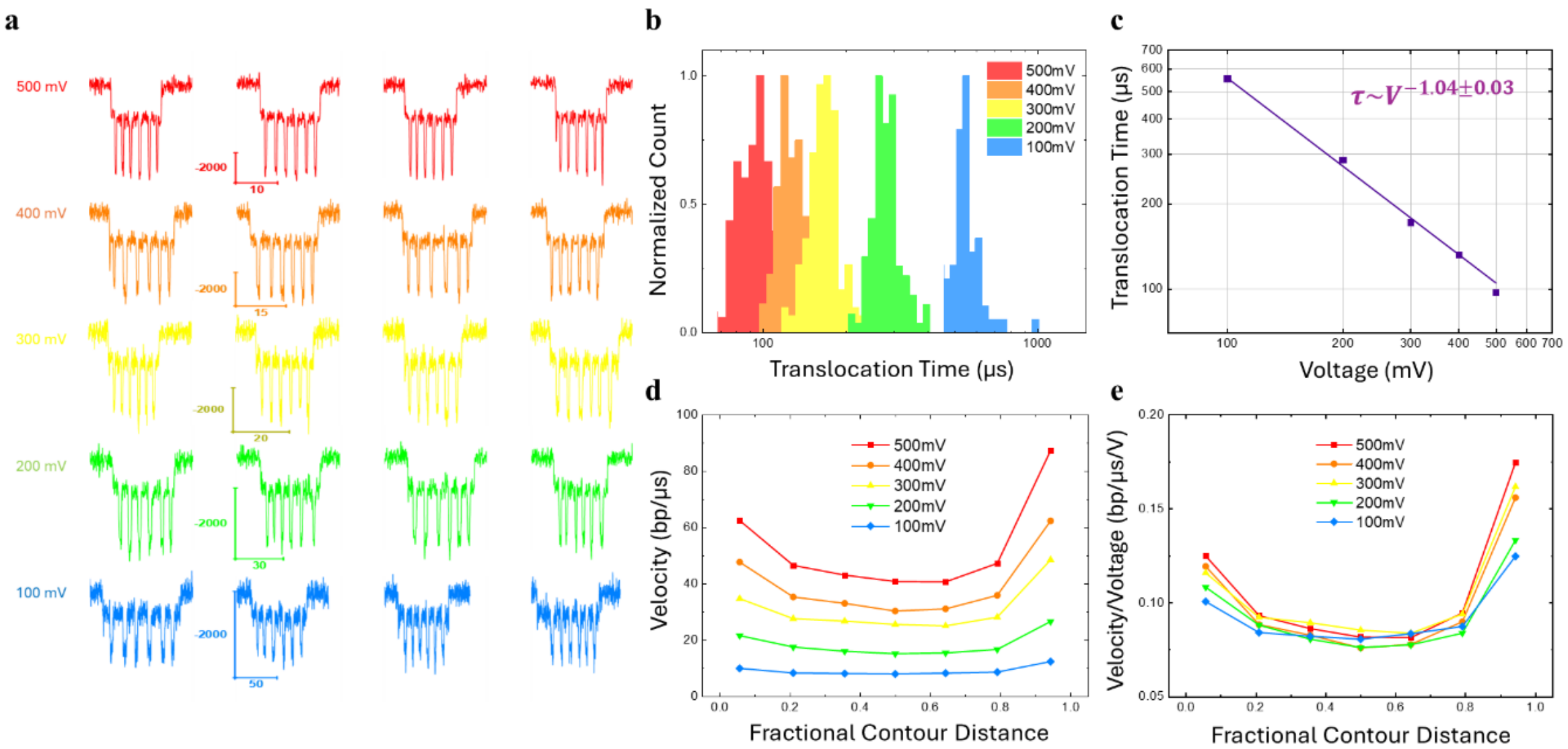

**Figure S13.** Effect of voltage on translocation velocity, measured in a 11 nm pore. **a)** Traces of VPM translocations under voltages ranging from 100 mV to 500 mV. **b)** Distribution of VPM translocation times in different voltages. **c)** Plot of mean translocation time vs voltage fitted to a power-scaling law $\tau \sim V^{-\beta}$ with $\beta = 1.04 \pm 0.03$. **d)** Translocation velocity profiles normalized by applied voltage.

Figure S13d plots the translocation velocity profiles measured for the different voltages. To normalize the effect of voltage resulting in velocities increasing five-fold in between the 100 and 500 mV measurements, Figure S13e instead plots the translocation velocities divided by the corresponding voltage. As a result, the velocities of the five non-extremity segments essentially overlap, as expected from a perfectly inverse voltage dependence. Interestingly, however, both the first and last DNA segment velocities show voltage dependence, with higher voltages resulting in higher velocities. At the moment of writing, it is unclear whether this observation arises from a physical phenomenon not fully understood yet, or simply due to the limitations of our analysis technique and its sensitivity to temporal resolution, as discussed in section S2 above. Regarding the latter, we note that the five-fold temporal difference between the 100 mV and 500 mV translocations is much larger than that of the pore-size and polymer length signals, which at maximum show a 2-fold change in velocity in going from a 9 nm to a 36 nm pore.